\documentclass[]{interact}

\usepackage{a4wide}                     %
\usepackage{amsmath}                    %
\usepackage[british]{babel}      %
\usepackage{fancyhdr}                   %
\usepackage{graphicx}                   %
\usepackage{grffile}                    %
\usepackage{pifont}                     %
\usepackage{paralist}                   %
\usepackage[usenames,dvipsnames]{color} %
\usepackage{moreverb}                   %
\usepackage{lscape}                     %
\usepackage{tocbibind}
\usepackage{verbatim}                   %
\usepackage{datetime}                   %
\usepackage{array}                      %
\usepackage[utf8]{inputenc}            %
\usepackage{ amssymb }                 %
\usepackage{amsbsy}
\usepackage{float}                     %
\usepackage{subcaption}
\usepackage{soul}

\usepackage{color}
\definecolor{darkred}{rgb}{0.5,0,0}
\definecolor{darkgreen}{rgb}{0,0.5,0}
\definecolor{darkblue}{rgb}{0,0,0.5}
\definecolor{gray}{rgb}{0.35,0.35,0.35}

\usepackage[pagebackref=false, %
            colorlinks=true,
            linkcolor=gray,
            bookmarks=true,
            filecolor=gray,
            urlcolor=gray,
            citecolor=gray
           ]{hyperref} %
\usepackage{lineno}                   %
\usepackage{longtable}                %
\usepackage[section]{placeins}        %
\usepackage{url}                      %
\usepackage{makeidx}                  %
\usepackage{hyperref}                 %
\usepackage{booktabs}                 %
\usepackage{rotating}                 %

\usepackage{todonotes}
\newcommand{\note}[1]{}

\newcommand{\rb}{\mbox{Rayleigh–B\'enard~}}

\newcommand{\Ra}{\mbox{Ra}}  %
\newcommand{\Pran}{\mbox{Pr}} %
\newcommand{\Nu}{\mbox{Nu}}  %

\begin{document}

\title{Controlling \rb convection via Reinforcement Learning}

\author{ \name{Gerben Beintema\textsuperscript{a},
     Alessandro Corbetta\textsuperscript{a},
    Luca Biferale\textsuperscript{b}, and Federico
    Toschi\textsuperscript{a,c,d}} \affil{
    \textsuperscript{a}Department of Applied Physics, Eindhoven
    University of Technology, 5600 MB Eindhoven, The Netherlands;
    \textsuperscript{b}Department of Physics and INFN, University of
    Rome Tor Vergata, I-00133 Rome, Italy;
    \textsuperscript{c}Department of Mathematics and Computer Science,
    Eindhoven University of Technology,  5600 MB;
    \textsuperscript{d}CNR-IAC, I-00185 Rome, Italy; } }

\maketitle

\begin{abstract}

  Thermal convection is ubiquitous in nature as well as in many
  industrial applications. The identification of effective control
  strategies to, e.g., suppress or enhance the convective
  heat exchange under fixed external thermal gradients is an
  outstanding fundamental and technological issue.  In this work, we
  explore a novel approach, based on a state-of-the-art Reinforcement
  Learning (RL) algorithm, which is capable of significantly reducing
  the heat transport in a two-dimensional Rayleigh-B\'enard system by
  applying small temperature fluctuations to the lower boundary of the
  system.  By using numerical simulations, we show that our RL-based
  control is able to stabilize the conductive regime and bring the
  onset of convection up to a Rayleigh number
  $Ra_c \approx 3 \cdot 10^4$, whereas in the uncontrolled case it
  holds $\Ra_{c}=1708$. Additionally, for $\Ra{} > 3 \cdot 10^4$, our
  approach outperforms other state-of-the-art control algorithms
  reducing the heat flux by a factor of about $2.5$. In the last part
  of the manuscript, we address theoretical limits connected to
  controlling an unstable and chaotic dynamics as the one considered
  here. We show that controllability is hindered by observability
  and/or capabilities of actuating actions, which can be quantified in
  terms of characteristic time delays. When these delays become comparable
  with the Lyapunov time of the system, control becomes impossible.
\end{abstract}


\section{Introduction}%

\rb Convection (RBC) provides a widely studied paradigm for
thermally-driven flows, which are ubiquitous in nature and in
industrial applications~\cite{intro:thermobook}.  Buoyancy effects,
ultimately yielding to fluid dynamics instability, are determined by
temperature gradients~\cite{intro:chandrasekhar-fluid-instalbility}
and impact on the heat transport. The control of RBC is an outstanding
research topic with fundamental scientific implications~\cite{intro:RB-control-accel-exp}.
Additionally, preventing, mitigating or enhancing such instabilities
and/or regulating the heat transport is crucial in numerous
applications. Examples include crystal growth
processes, e.g. to produce silicon wafers \cite{intro:crystal-convection-inhomogeneities-book}. 
Indeed, while the speed of these processes benefits from increased
temperature gradients, the quality of the outcome is endangered by
fluid motion (i.e. flow instability) that grows as the thermal
gradients increase. Thus the key problem addressed here: can we
control and stabilize fluid flows that, due to temperature gradients,
would otherwise be unstable?

\noindent In the Boussinesq
approximation, the fluid motion in RBC can be described via the following equations~\cite{intro:lbm-book}:  %
\begin{align}
 &\frac{\partial \mathbf{u}}{\partial t} + \mathbf{u} \cdot
\pmb{\nabla}\mathbf{u}  = -\pmb{\nabla} p + \nu \nabla^2
                             \mathbf{u} + \mathbf{\hat{y}} \alpha g (T-T_0), \\
   &\frac{\partial T}{\partial t} + \mathbf{u} \cdot
\pmb{\nabla}T = \kappa \nabla^2 T.
\end{align}
where $t$ denotes the time, $\mathbf{u}$ the incompressible velocity
field ($\pmb{\nabla} \cdot \mathbf{u} = 0$), $p$ the pressure, $\nu$
the kinematic viscosity, $\alpha$ the thermal expansion coefficient,
$g$ the magnitude of the acceleration of gravity (with direction
$\mathbf{\hat{y}}$), and $\kappa$ the thermal diffusivity. For a
fluid system confined between two parallel horizontal planes at
distance $H$ and with temperatures $T_C$ and $T_H = T_C + \Delta T$,
respectively for the top and the bottom element ($\Delta T > 0$), it
is well known that the dynamics is regulated by three non-dimensional
parameters: the Rayleigh and Prandtl numbers and the aspect ratio of
the cell (i.e. the ratio between the cell height and width $L_x$), i.e.
\begin{equation}
  \Ra{} = \frac{g \alpha (T_H - T_C) H^3}{\kappa \nu}, \qquad \Pran = {\nu}/{\kappa}, \qquad \Gamma = {H}/{L_x}.
\end{equation}
Considering adiabatic side walls, a mean heat flux, $q$, independent on the
height establishes on the cell:
\begin{equation}
 q =  \overline{\langle u_y T \rangle}_{x}  -\kappa \partial_y \overline{\langle T  \rangle}_{x} = \text{const},
 \label{eq:heat-flux}
\end{equation}
where $\langle \cdot \rangle_x$ indicates an average with respect to
the $x$-axis, parallel to the plates, and  $\overline{\bullet}$ the time averaging. The time-averaged heat flux is customarily
reported in a non-dimensional form, scaling it by the conductive heat
flux, $\kappa \Delta T/H$, which defines the Nusselt number
\begin{equation}
  \Nu = \frac{q}{\kappa \Delta T/H}.
  \label{eq:averageNu}
\end{equation}
As the Rayleigh number overcomes a critical threshold, $\Ra{}_c$, fluid
motion is triggered enhancing the heat exchange ($\Nu > 1$).\\

\begin{figure}[t!]
\centering
\begin{subfigure}{0.461290323\textwidth}
  \centering
  \includegraphics[width=1.0\linewidth]{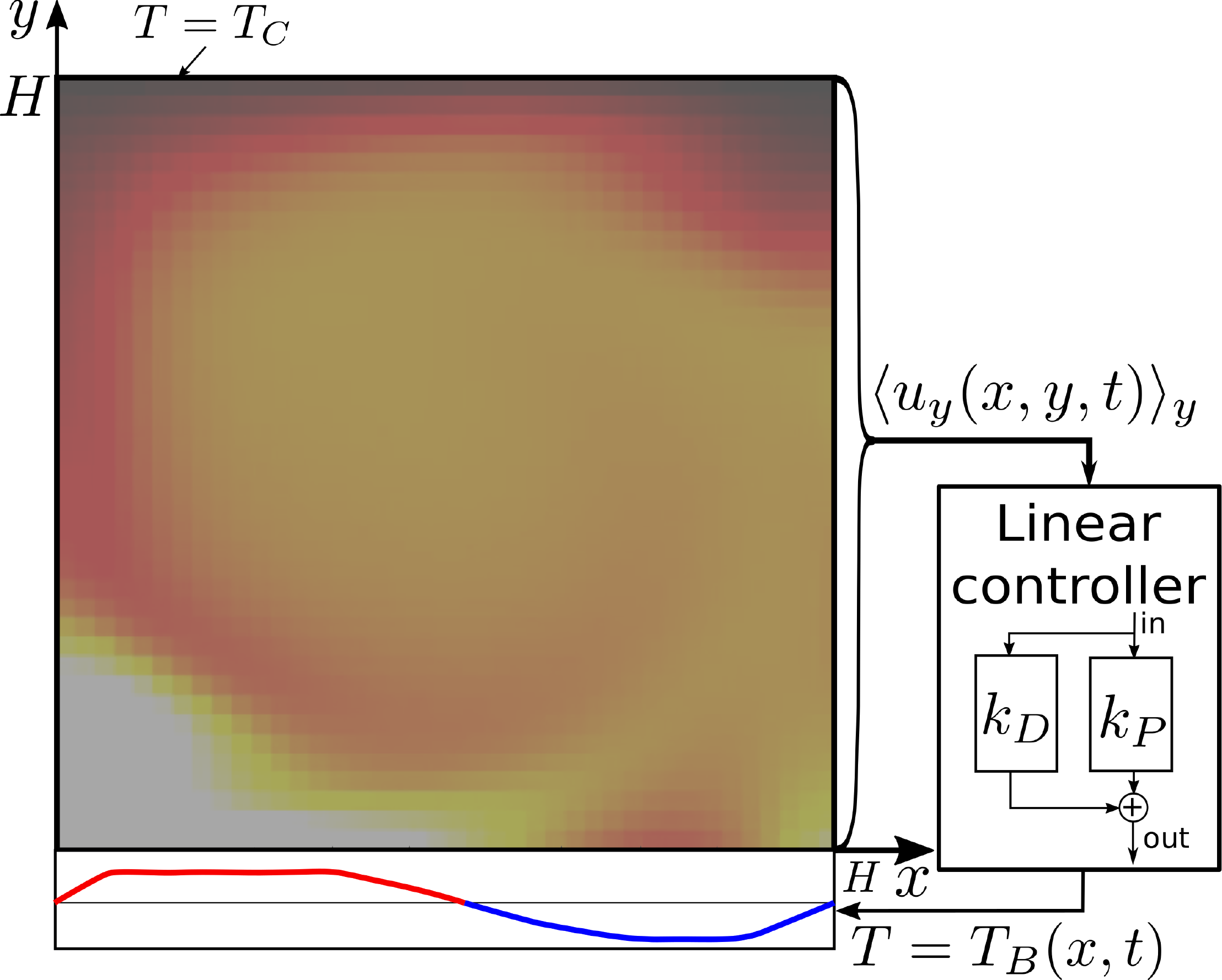}
  \caption{Linear Control}
  \label{fig:lin-scheme}
\end{subfigure}%
\begin{subfigure}{0.538709677\textwidth}
  \centering
  \includegraphics[width=1.0\linewidth]{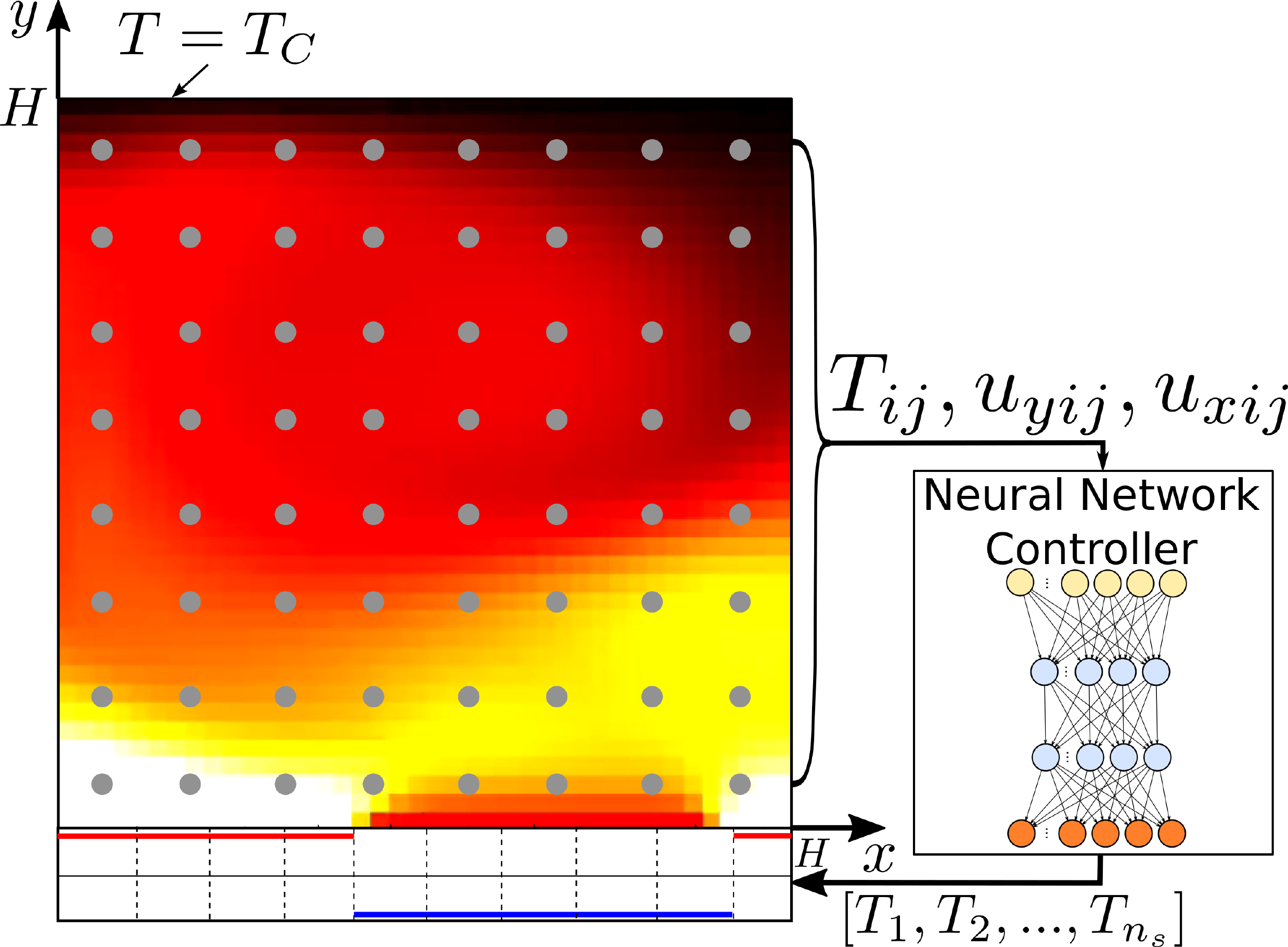}
  \caption{RL control}
  \label{fig:RL-scheme}
\end{subfigure}
\caption{Schematics of linear (a) and Reinforcement Learning (b)
  control methods applied to a \rb system and aiming at reducing
  convective effects (i.e. the Nusselt number). The system consists of
  a domain with height, $H$, aspect ratio, $\Gamma=1$, no-slip
  boundary conditions, constant temperature $T_C$ on the top boundary,
  adiabatic side boundaries and a controllable bottom boundary where
  the imposed temperature $T_B(x,t)$ can vary in space and time
  (according to the control protocol) while keeping a constant average
  $\langle T_B(x,t) \rangle_x = T_H$. Because the average temperature
  of the bottom plate is constant, the Rayleigh number is well-defined
  and constant over time. The control protocol of the linear
  controller (a) works by calculating the distance
  measure $E(x,t)$ from the ideal state
  (cf. Eq.~\eqref{eq:pd-contr}-\eqref{eq:metric-linear}) and, based on
  linear relations, applies temperature corrections to the bottom
  plate.
  The RL method (\ref{fig:RL-scheme}) uses a Neural Network controller
  which acquires flow state from a number of probes at fixed locations
  and returns a temperature profile (see details in
  Fig.~\ref{fig:NN_expanded}). The parameters of the Neural Network
  are automatically optimized by the RL method, during
  training. Moreover, in the RL case, the temperature fluctuations at
  the bottom plate are piece-wise constant and can have only prefixed
  temperature value between two states, hot or cold
  (cf. Eq.~\eqref{eq:tkpieces}).  Operating with discrete actions,
  reduces significantly the complexity and the computational resources
  needed for training.}
\label{fig:exp-setup}
\end{figure}

Dubbed in terms of Rayleigh and Nusselt numbers, our control question
becomes: can we diminish, or minimize, Nusselt for a fixed Rayleigh
number?

\noindent In recent years, diverse approaches have been proposed to
tackle this issue. These can be divided into passive and active
control methods. Passive control methods include: acceleration
modulation~\cite{intro:RB-control-accel-exp,intro:RB-control-accel-compute},
oscillating shear flows~\cite{intro:RB-control-shear-flows}, and
oscillating boundary
temperatures~\cite{intro:RB-control-temperature-oscillations}. Active
control methods include: velocity actuators
\cite{intro:RB-control-velocity}, and perturbations of the thermal
boundary layer \cite{intro:RB-control-temp-first,
  intro:RB-control-temp-experimental,intro:RB-control-temp-robust}.
Many of these methods, although ingenious, are not practical due,
e.g., to the requirement of a perfect knowledge of the state of the
system, or a starting condition close to the conductive state -
something difficult to
establish~\cite{intro:RB-control-temp-experimental}.  Another
state-of-the-art active method, that is used in this paper as
comparison, is based on a linear control acting at the bottom thermal
boundary layer~\cite{intro:PDRBcontrol}.
The main difficulty in controlling RBC resides in its chaotic behavior
and chaotic response to controlling actions. In recent years,
Reinforcement Learning (RL) algorithms~\cite{intro:RL-book} have been
proven capable of solving complex control problems, dominating
extremely hard challenges in high-dimensional spaces (e.g. board
games~\cite{intro:RL-example-GO,intro:RL-example-boardgames} and
robotics~\cite{intro:RL-example-irl-cube}).
Reinforcement Learning is a supervised Machine Learning
(ML)~\cite{RLback:ML-book} approach aimed at finding optimal control
strategies. This is achieved by successive trial-and-error
interactions with a (simulated or real) environment which iteratively
improve an initial random control policy. Indeed, this is usually a
rather slow process which may take millions of trial-and-error
episodes to converge~\cite{intro:RL-example-rainbow-atari}.
Reinforcement learning has also been used in connection with fluid
flows, such as for training smart inertial or self-propelling
particles~\cite{colabrese2018smart,colabrese2017flow,gustavsson2017finding},
for schooling of fishes~\cite{gazzola2016learning,verma2018efficient},
soaring of birds and gliders in turbulent
environments~\cite{reddy2016learning,reddy2018glider}, optimal
navigation in turbulent flows~\cite{biferale2019,alageshan2019path},
drag reduction by active control in the turbulent
regime~\cite{intro:drag}, and
more~\cite{muinos2018reinforcement,novati2018deep,tsang2018self,intro:drag-accel,cichos2020machine,rabault2020deep}.

In this work, we show that RL methods can be successfully applied for
controlling a \rb system at fixed Rayleigh number reducing (or
suppressing) convective effects. Considering a 2D proof-of-concept
setup, we demonstrate that RL can significantly outperform
state-of-the-art linear methods~\cite{intro:PDRBcontrol} when allowed
to apply (small) temperature fluctuations at the bottom plate (see
setup in Figure~\ref{fig:exp-setup}). In particular, we target a
minimization of the time-averaged Nusselt number
(Eq.~\eqref{eq:averageNu}), aiming at reducing its instantaneous
counterpart:
\begin{equation}
    \Nu_{\text{instant}}(t) =  \frac{\langle u_y T \rangle_{x,y}  -\kappa \partial_y \langle T  \rangle_{x,y}}{\kappa \Delta T/H},
    \label{eq:Nu_instant}
  \end{equation}
where the additional average along the vertical direction, $y$, amends
instantaneous fluctuations.

Finding controls fully stabilizing RBC might be, however, at all
impossible. For a chaotic system as RBC, this may happen, among
others, when delays in controls or observation become comparable with
the Lyapunov time. We discuss this topic in the last part of the paper
employing Reinforcement Learning to control the Lorenz attractor, a
well-known, reduced version of RBC~\cite{back:lorenz1963}.

The rest of this manuscript is organized as
follows. %
In Section~\ref{sec:rb-control-background-and-implementation}, we
formalize the \rb control problem and the implementation of both
linear and RL controls. In Section~\ref{sec:results}, we present the
control results and comment on the induced flow structures. In
Section~\ref{sec:limits} we analyze physical factors that
limit the RL control performance. The discussion in
Section~\ref{sec:conclusion} closes the paper.

\section{Control-based convection reduction} %
\label{sec:rb-control-background-and-implementation}

In this section we provide details of the \rb system considered, formalize the control problem and introduce the control methods.

We consider an ideal gas ($\Pran=0.71$) in a two-dimensional \rb
system with $\Gamma = 1$ at an assigned Rayleigh number (cf. sketch
and $(x,y)$ coordinate system in Figure~\ref{fig:exp-setup}). We
assume the four cell sides to satisfy a non-slip boundary condition,
the lateral sides to be adiabatic, and a uniform temperature, $T_C$,
imposed at the top boundary. We enforce the Rayleigh number by
specifying the average temperature,
\begin{equation}
    T_H = \langle T_B(x,t)\rangle_x,
\label{eq:control-constraint-average}
\end{equation}
at the bottom boundary (where
$T_B(x,t)$ is the instantaneous temperature at location $x$ of the
bottom boundary). Temperature fluctuations with respect to such
average,
\begin{equation}
    \hat {T}_B(x,t) = T_B(x,t) - T_H,
\end{equation}
are left for control.

We aim at controlling $\hat {T}_B(x,t)$ to minimize convective
effects, which we quantify via the time-averaged Nusselt number
(cf. Eq.\eqref{eq:averageNu}).
We further constrain the allowed temperature fluctuations to
\begin{equation}
|\hat{T}_B(x,t)| \leq C \qquad \forall x,t,
\label{eq:bound-temp-fluct}
\end{equation}
to prevent extreme and nonphysical temperature gradients (in similar spirit to the experiments in~\cite{intro:RB-control-temp-experimental}).

We simulate the flow dynamics through the Lattice-Boltzmann method
(LBM)~\cite{intro:lbm-book} employing a double lattice, respectively
for the velocity and for the temperature populations (with D2Q9 and
D2Q4 schemes on a square lattice with sizes $N_x=N_y$; collisions are
resolved via the standard BGK relaxation).
We opt for the LBM since it allows for fast, extremely vectorizable,
implementations which enables us to perform multiple (up to hundreds)
simulations concurrently on a GPU architecture. See
Table~\ref{tab:sim-params} for relevant simulation parameters; further
implementation details are reported in
Appendix~\ref{sec:appendix-RB-sim}.

\begin{table}[ht]   
\centering
\caption{\rb system and simulation parameters.} %
{\renewcommand{\arraystretch}{1.1}
\begin{tabular}{lll}
\hline
\multicolumn{3}{c}{Parameters}                                                                                                       \\ \hline
$\Ra{}$      & \multicolumn{1}{l|}{$10^3 \xrightarrow{} 10^7$}              & Rayleigh Number                                          \\
$\Pran$    & \multicolumn{1}{l|}{$0.71$}                                  & Prandtl Number                                           \\
$\Gamma$   & \multicolumn{1}{l|}{$1$}                                     & Aspect Ratio                                             \\ \hline
\multicolumn{3}{c}{Control}                                                                                                          \\ \hline
$C$        & \multicolumn{1}{l|}{0.75}                                    & Control amplitude limit,  Eq.~\eqref{eq:bound-temp-fluct} \\
$\Delta t$ & \multicolumn{1}{l|}{$16 \xrightarrow{} 180$}                 & Control loop (unit: LBM steps)                           \\
$t_i$      & \multicolumn{1}{l|}{$0$  (training)}                         & Start evaluation time, for averages in Eq.~\eqref{eq:averageNu}       \\
"        & \multicolumn{1}{l|}{$150 \Delta t$ (test)}                   & \multicolumn{1}{c}{"}                                   \\
$t_e$      & \multicolumn{1}{l|}{$500 \Delta t$}                          & End evaluation time, for averages in Eq.~\eqref{eq:averageNu}         \\ \hline
\multicolumn{3}{c}{Lattice Boltzmann simulation (Appendix~\ref{sec:appendix-RB-sim})}                                                \\ \hline
$N_X=N_Y$  & \multicolumn{1}{l|}{$20 \xrightarrow{} 350$}                 & Grid size                                                \\
$c_s^2$    & \multicolumn{1}{l|}{$1/3$}                                   & Speed of sound                                           \\
$\tau$     & \multicolumn{1}{l|}{0.56}                                    & Relaxation time                                          \\
$T_C$      & \multicolumn{1}{l|}{1}                                       & Top boundary  temperature                                \\
$T_H$      & \multicolumn{1}{l|}{2}                                       & Bottom boundary mean temperature                         \\
$\tau_T$   & \multicolumn{1}{l|}{$(\tau - 1/2)/\Pran + 1/2$}              & Temperature relaxation time                              \\
$\nu$      & \multicolumn{1}{l|}{$c_s^2 (\tau - 1/2)$}                    & Kinematic viscosity                             \\
$\kappa$   & \multicolumn{1}{l|}{$1/4 (2 \tau_T - 1)$}                    & Thermal diffusivity                                      \\
$\alpha g$ & \multicolumn{1}{l|}{$\Ra{} \frac{\kappa \nu}{(T_H-T_C) NY^3}$} & Effective gravity                                        \\ \hline
\end{tabular}
}
\label{tab:sim-params}
\end{table}

Starting from a system in a (random) natural convective state (cf. experiments~\cite{intro:RB-control-temp-experimental}), 
controlling actions act continuously. Yet, to limit learning
computational costs, $\hat T_B(x,t)$ is updated with a period,
$\Delta t$ (i.e. control loop), longer than the LBM simulation step
and scaling with the convection time,
$t_{\text{convection}} \sim H/U_{\text{bulk}}$. We report in
Table~\ref{tab:RB-envs} the loop length, which satisfies,
approximately, $\Delta t \approx 1/20\, t_{\text{convection}}$, and
the system size, all of which are \Ra{}-dependent. Once more, for
computational efficiency reasons, we retain the minimal system size
that enables us to quantify the (uncontrolled) Nusselt number within
$5\%$ error (Appendix~\ref{sec:appendix-RB-sim}).

In the next subsections we provide details on the linear and
reinforcement-learning based controls.

\subsection{Linear control}

We build our reference control approach via a linear
proportional-derivative (PD) controller~\cite{intro:PDRBcontrol}. Our
control loop prescribes instantaneously the temperature fluctuations
at the bottom boundary as
\begin{equation}
  \hat T_B(x,t) = \mathcal{R}( \tilde T_B(x,t) ) = - \mathcal{R}( (k_P - k_D \partial_t)E(x,t) )
  \label{eq:pd-contr}
\end{equation}
with $k_P$, $k_D$ being constant parameters and the (signed)
distance from the optimal conductive state ($u_x=u_y=0$), $E(x,t)$, satisfying 
\begin{equation}
  E(x,t) = \langle u_y(x,y,t) \rangle_y/V_0,  
\label{eq:metric-linear}
\end{equation}
where $V_0$ is a reference vertical velocity. To ensure the constraints given by Eqs.~\eqref{eq:control-constraint-average},\eqref{eq:bound-temp-fluct} we operate a clipping and renormalization operation, $\mathcal{R}(\cdot)$, as described in Appendix~\ref{sec:app-normaliz}.

Various other metrics, $E(x,t)$, have been proposed leveraging, for
instance, on the shadow graph method
($E(x,t) = (\langle \rho(x,y,t) \rangle_y - \rho_0)/\rho_0$
\cite{intro:RB-control-temp-experimental}), and on the mid-line
temperature ($E(x,t) = (T(x,H/2,t) - T_{1/2})/\Delta T$, with
$T_{1/2} = 1/2(T_H + T_C)$, \cite{intro:RB-control-temp-first}). These
metrics provide similar results and, in particular, an average Nusselt
number for the controlled systems within $5\%$. Hence, we opted for
Eq.~\eqref{eq:metric-linear} as it proved to be more stable. Note
that, by restricting to $k_D=0$, one obtains a linear proportional (P)
controller. While supposedly less powerful than a PD controller, in
our case the two show similar performance. The controller operates
with the same space and time discretization of the LBM simulations,
with the time derivative of $E(x,t)$ calculated with a first order
backwards scheme.  We identify the Rayleigh-dependent parameters $k_P$
and $k_D$ via a grid-search algorithm~\cite{RLback:optimization-book}
for $\Ra{} \leq 10^6$ (cf. values in
Table~\ref{tab:PD-parameters}). In case of higher $\Ra{}$, due to the
chaoticity of the system, we were unable to find parameters
consistently reducing the heat flux with respect to the uncontrolled
case.

\begin{table}[ht]   
\centering
\caption{Parameters used for the linear controls in Lattice Boltzmann units (cf. Eq.~\eqref{eq:pd-contr}; note: a P controller is obtained by setting $k_D = 0$ ). At $\Ra{}=10^7$ we were unable to find PD controllers performing better than P controllers, which were anyway ineffective.} %
\begin{tabular}{rr|r|rr}
  \hline\\
\multicolumn{1}{c}{}   & \multicolumn{1}{c|}{}            & \multicolumn{1}{c|}{P control} & \multicolumn{2}{c}{PD control}                        \\ \hline
\multicolumn{1}{l}{Ra} & \multicolumn{1}{l|}{$N_x = N_y$} & \multicolumn{1}{c|}{$k_P$}     & \multicolumn{1}{c}{$k_P$} & \multicolumn{1}{c}{$k_D$} \\ \hline
$1 \cdot 10^3$         & 20                               & 0.0                            & 0.0                       & 0.0                       \\
$3 \cdot 10^3$         & 20                               & $3.16 \cdot 10^2$              & $3.16 \cdot 10^2$         & 0.0                       \\
$1 \cdot 10^4$         & 20                               & $4.12 \cdot 10^2$              & $5.28 \cdot 10^2$         & $8.24 \cdot 10^4$         \\
$3 \cdot 10^4$         & 25                               & $8.97 \cdot 10^2$              & $5.45 \cdot 10^4$         & $1.91 \cdot 10^6$         \\
$1 \cdot 10^5$         & 30                               & 16.4                           & 94.8                      & $1.05 \cdot 10^4$         \\
$3 \cdot 10^5$         & 40                               & 9.38                           & 11.5                      & $1.87\cdot 10^3$          \\
$1 \cdot 10^6$         & 100                              & 6.61                           & $1.84 \cdot 10^4$         & $3.06 \cdot 10^5$         \\
$3 \cdot 10^6$         & 200                              & 7.38                           & 0.12                      & 31.8                      \\
$1 \cdot 10^7$         & 350                              & 0.33                           & -                         & -                         \\ \hline
\end{tabular}
\label{tab:PD-parameters}
\end{table}

\subsection{Reinforcement Learning-based control}

In a Reinforcement Learning context we have a policy, $\pi$, that
selects a temperature fluctuation, $\hat {T}(x,t)$, on the basis of
the observed system state. $\pi$ is identified automatically through
an optimization process, which aims at maximizing a reward signal. In
our case we define the system state, the allowed controlling actions
and the reward are as follows:
\begin{itemize}
\item The \textit{state space}, $S$, includes observations of the
  temperature and velocity fields (i.e. of $n_f = 3$ scalar fields)
  probed on a regular grid in $G_X \times G_Y = 8 \times 8$ nodes for
  the last $D=4$ time steps (i.e.
  $t, t -\Delta t, \ldots, t - (D-1) \Delta t$, where $t$ is the
  current time). Note that the probe grid has a coarser resolution
  than the lattice, i.e. $G_X< N_X$, $G_Y< N_Y$, which allows to
  reduce the complexity of the control problem. It holds, therefore,
  $S = \mathbb{R}^{D\cdot n_f \cdot G_X \cdot G_Y}$.
\item The \textit{action space}, $A$, includes the temperature
  profiles for the lower boundary that the controller can
  instantaneously enforce. To limit the computational
  complexity, we allow profiles which are piece-wise constant
  functions on $n_s = 10$ segments
  (cf. Figure~\ref{fig:exp-setup}). Moreover, each of the $n_s$
  function elements, $\tilde T_k$ ($k=1,2,\ldots,n_s$), can  attain only two temperature
  levels, i.e.
  \begin{equation}
      \tilde T_k \in \{C,-C\}.
      \label{eq:tkpieces}
  \end{equation}
  To enforce the constraint in
  Eq.~\eqref{eq:control-constraint-average},\eqref{eq:bound-temp-fluct}
  we normalize the profile according to
  Appendix~\ref{sec:app-normaliz}, generating the final profile
  $\hat{T}_k=\mathcal{R}(\tilde{T}_k)$.
    After normalization, the action space includes
     $2^{n_s}-1$ distinct actions.
  
  \item The \textit{reward function} defines the goal of the control
    problem. We define the
    reward, $r_{l+1}$, as the negative instantaneous Nusselt number (cf. Eq.~\ref{eq:Nu_instant}) which results
    from applying a temperature profile $a_l \in A$ at time
    $t_l$. In formulas, it holds
    \begin{equation}
    r_{l+1} = -\Nu_{\text{instant}}(t_{l+1}). %
    \end{equation}
    Note that the RL controller aims at maximizing the reward accumulated over time (rather than the instantaneous reward), which minimizes the average Nusselt number, Eq.~\eqref{eq:averageNu}, as desired.
\end{itemize}
We employ the Proximal Policy Optimization (PPO) RL
algorithm~\cite{intro:PPO-algorithm}, which belongs to the family of
Policy Gradient Methods. Starting from a random initial condition,
Policy Gradient Methods search iteratively (and probabilistically) for
optimal (or sufficient) policies by gradient ascent (based on local
estimates of the performance). Specifically, this optimization employs
a probability distribution, $\pi(a_i|s_i)$, on the action space
conditioned to the instantaneous system state. At each step of the
control loop, we sample and apply an action according to the
distribution $\pi(a|s)$. Notably, the sampling operation is
  essential at training time, to ensure an adequate balance between
  exploration and exploitation, while at test time, this stochastic
  approach can be turned deterministic by restricting to the action
  with highest associated probability. In our case, at test time we
  used the deterministic approach for $\Ra{} < 3 \cdot 10^6$ and the
  stochastic approach for $\Ra{} \geq 3 \cdot 10^6$, as this allowed for
  higher performance.

  The PPO algorithm is model-free, i.e. it does not need assumptions
  on the nature of the control problem. Besides, it does not generally
  require significant hyperparameter tuning, as often happens for RL
  algorithms (e.g. value based method~\cite{intro:PPO-algorithm}).

When the state vector $s_i$ is high-dimensional (or even continuous),
it is common to parameterize the policy function in probabilistic
terms as $\pi(a_i|s_i) = \pi(a_i|s_i;\boldsymbol{\theta})$, for a
parameter vector $\boldsymbol{\theta}$~\cite{intro:RL-example-GO}.
This parameterization can be done via different kinds of functions
and, currently, neural networks are a popular
choice~\cite{intro:universal-NN}. In the simplest case, as used here,
the neural network is a multilayer perceptron~\cite{intro:RL-book}
(MLP). An MPL is a fully connected network in which neurons are
stacked in layers and information flows in a pre-defined direction
from the input to the output neurons via ``hidden'' neuron layers. The
$i$-th neuron in the $(n+1)$-th layer operates returning the value
$h^{(n+1)}_i$, which satisfies
\begin{equation}
    h^{(n+1)}_i = \sigma\left(b_i^{(n)} + \sum_j A_{ij}^{(n)} h^{(n)}_j \right),
\end{equation}
where the $h^{(n)}_j$'s are the outputs of the neurons of the previous
layer (the $n$-th one), which thus undergo an affine transformation
via the matrix $A_{ij}^{(n)}$ and the biases $b_i^{(n)}$. The non-linear
activation function $\sigma$ provides the network with the capability
of approximating non-linear
functions~\cite{intro:universal-NN}. During training, the parameters
$\boldsymbol{\theta}$ get updated through back
propagation~\cite{intro:back-prop} (according to the loss defined by
the PPO algorithm) which results in progressive improvements of the
policy function.

\begin{figure}[ht]
\setlength{\unitlength}{1cm}
\centering
\includegraphics[width=0.8\linewidth]{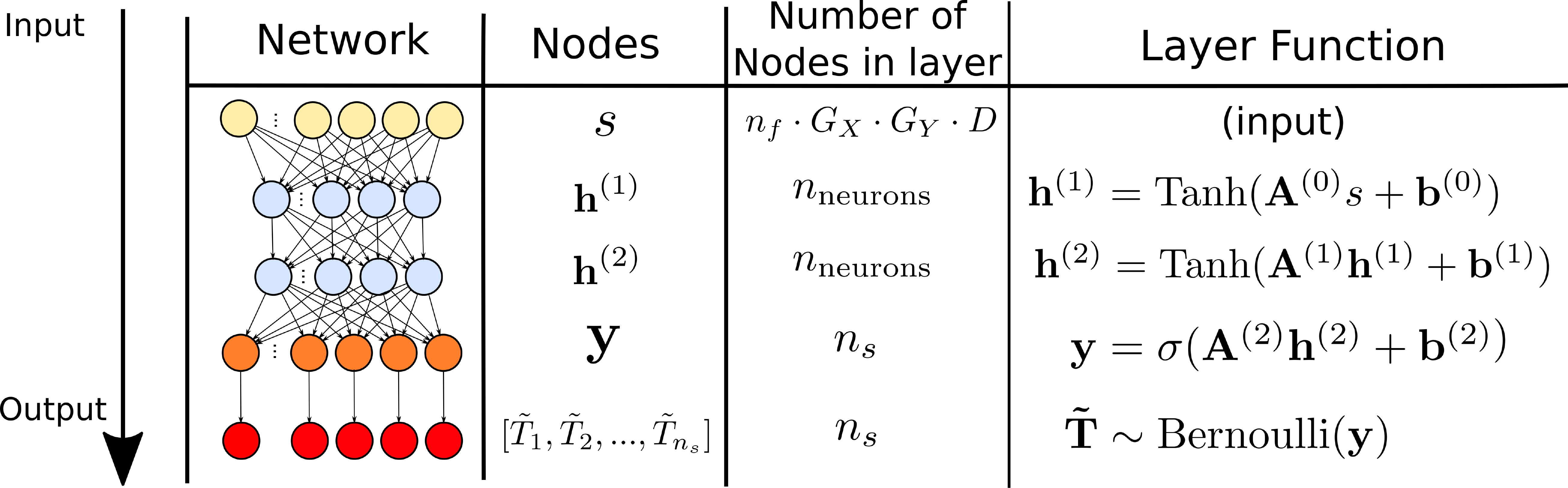}
\caption{Sketch of the neural network determining the policy adopted
  by the RL-based controller (policy network, $\pi$,
  cf. Fig.~\ref{fig:exp-setup}(b) for the complete setup). The input
  of the policy network is a state vector, $\textbf{s}$, which stacks
  the values of temperature and both velocity components for the
  current and the previous $D-1 = 3$ timesteps. Temperature and
  velocity are read on an evenly spaced grid of size $G_X=8$ by
  $G_Y=8$. Hence, $\textbf{s}$ has dimension
  $n_f \cdot G_X \cdot G_Y \cdot D = 3 \cdot 8 \cdot 8 \cdot 4 =
  768$. The policy network $\pi$ is composed of two fully connected
  feed forward layers with $n_{\text{neurons}} = 64$ neurons and
  $\sigma(\cdot) = \mbox{tanh}(\cdot)$ activation. The network output
  is provided by one fully connected layer with
  $\sigma(\cdot) = \phi(\cdot)$ activation
  (Eq.~\eqref{eq:sigmoid}). This returns the probability vector
  $\mathbf{y}=[y_1,y_2,\ldots,y_{n_s}]$. The $kj$-th bottom segment
  has temperature $C$ with probability $y_k$
  (Eq.~\eqref{eq:Bernoulli}).  This probability distribution gets
  sampled to produce a proposed temperature profile
  $\tilde{\textbf{T}} = (\tilde {T}_1,\tilde {T}_2,\ldots,\tilde
  {T}_{n_s})$. The final temperature fluctuations
  $\hat{T}_1,\hat{T}_2,\ldots,\hat{T}_{n_s}$ are generated with the
  normalization step in Eq.~\eqref{eq:normalization-temp}
  (cf. Eqs.~\eqref{eq:control-constraint-average}
  and~\eqref{eq:bound-temp-fluct}). The
  temperature profile obtained is applied to the bottom plate during a
  time interval $\Delta t$ (control loop), after which the procedure is
  repeated. } \label{fig:NN_expanded}
\end{figure}

To increase the computational efficiency, we opt for a policy function factorized as follows
\begin{equation}
    \pi(a_i | s_i) = \pi \left ( \tilde T_1, \tilde T_2,\ldots, \tilde T_{n_s} \middle | s_i \right ) =
    \prod_{k=1}^{n_s} \pi_k \left ( \tilde T_k \middle | s_i \right ).
\end{equation}
In other words, we address the marginal distributions of the local
temperature values $\tilde T_1, \tilde T_2,\ldots, \tilde T_{n_s}$.
We generate the marginals by an MLP (with two hidden layers each with
$\sigma(\cdot)=\mbox{Tanh}(\cdot)$ activation) that has $n_s$ final
outputs, $y_1,y_2,\dots,y_{n_s}$, returned by sigmoid activation
functions, in formulas:
\begin{equation} \label{eq:sigmoid}
y_k = \phi(z_k) = \frac{1}{1 + \exp(-z_k)},\qquad k=1,2,\ldots,n_s 
\end{equation}
with $z_k = (\sum_{j}A^{(2)}_{kj} h_j^{(2)} + b_k^{(2)}$ (see
Fig.~\ref{fig:NN_expanded}, note that $0\leq \phi(z_k) \leq 1$).  The
values $y_1,y_2,\dots,y_{n_s}$ provide the parameters for $n_s$
Bernoulli distributions that determine, at random, the binary
selection between the temperature levels $\{-C,C\}$. In formulas, it
holds
\begin{equation} \label{eq:Bernoulli}
    \pi_k \left ( \tilde T_k = + C \middle | s_i \right ) = \text{Bernoulli}(p=y_k).
\end{equation}
The final temperature profile is then determined via
Eq.~\eqref{eq:tkpieces} and the normalization in
Eq.~\eqref{eq:normalization-temp}.  We refer to
Fig.~\ref{fig:NN_expanded} for further details on the network.

\begin{figure}[t]
\setlength{\unitlength}{1cm}
\centering
\includegraphics[width=13cm]{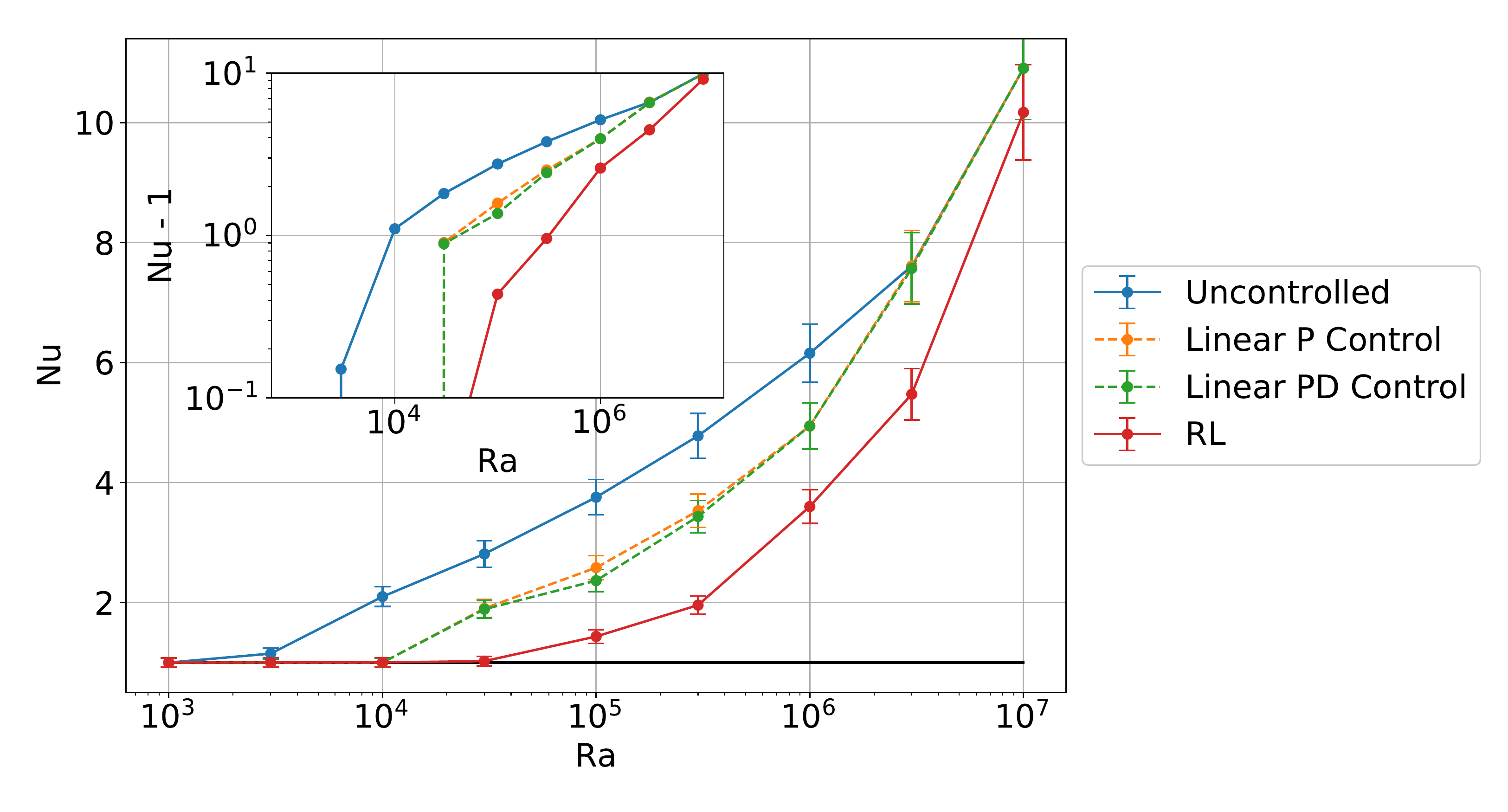}
\caption{Comparison of Nusselt number, averaged over time and ensemble,
  measured for uncontrolled and controlled \rb systems (note that all
  the systems are initialized in a natural convective state). We
  observe that the critical Rayleigh number, $\Ra{}_c$, increases when we
  control the system, with $\Ra{}_c = 10^4$ in case of linear control
  and $\Ra_c = 3\cdot 10^4$ in case of the RL-based control.
  Furthermore, for $\Ra> \Ra_c$, the RL control achieves a Nusselt number consistently
  lower than what measured in case of the uncontrolled system and for
  linear controls  (P and PD
    controllers have comparable performance at all the considered
    Rayleigh numbers, see also~\cite{intro:PDRBcontrol}). The error bars are estimated
  as $\mu_{\text{Nu}}/\sqrt{N}$, where $N=161$ is the number of
  statistically independent samplings of the Nusselt number.
}\label{fig:Ra-Nu-Main}%
\end{figure}

\section{Results} \label{sec:results} %

\begin{figure}[h]
\centering
\begin{subfigure}{.5\textwidth}
  \centering
  \includegraphics[width=1.0\linewidth]{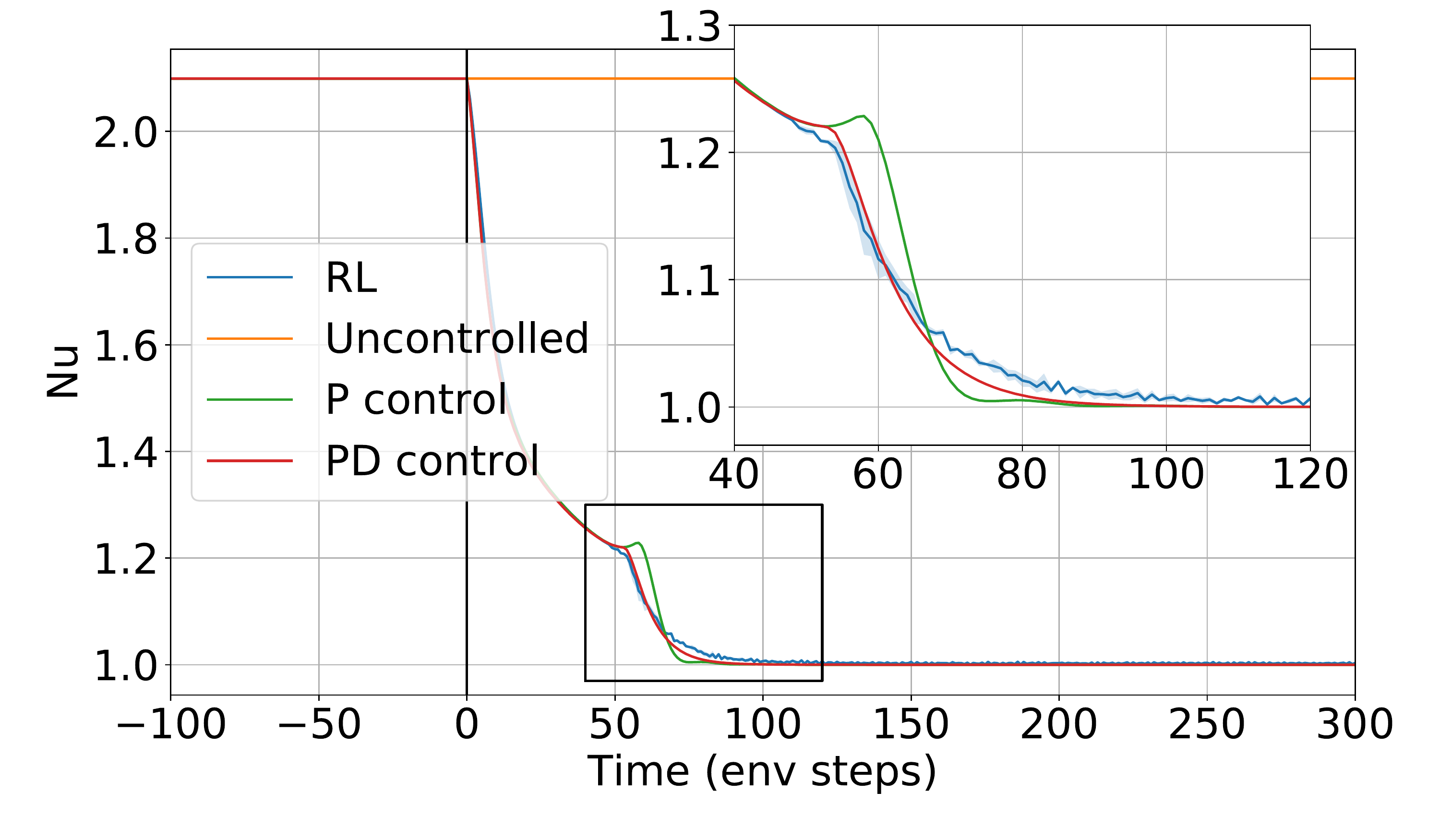}
  \caption{$\text{Ra} = 1 \cdot 10^4$}
  \label{fig:time-1e4}
\end{subfigure}%
\begin{subfigure}{.5\textwidth}
  \centering
  \includegraphics[width=1.0\linewidth]{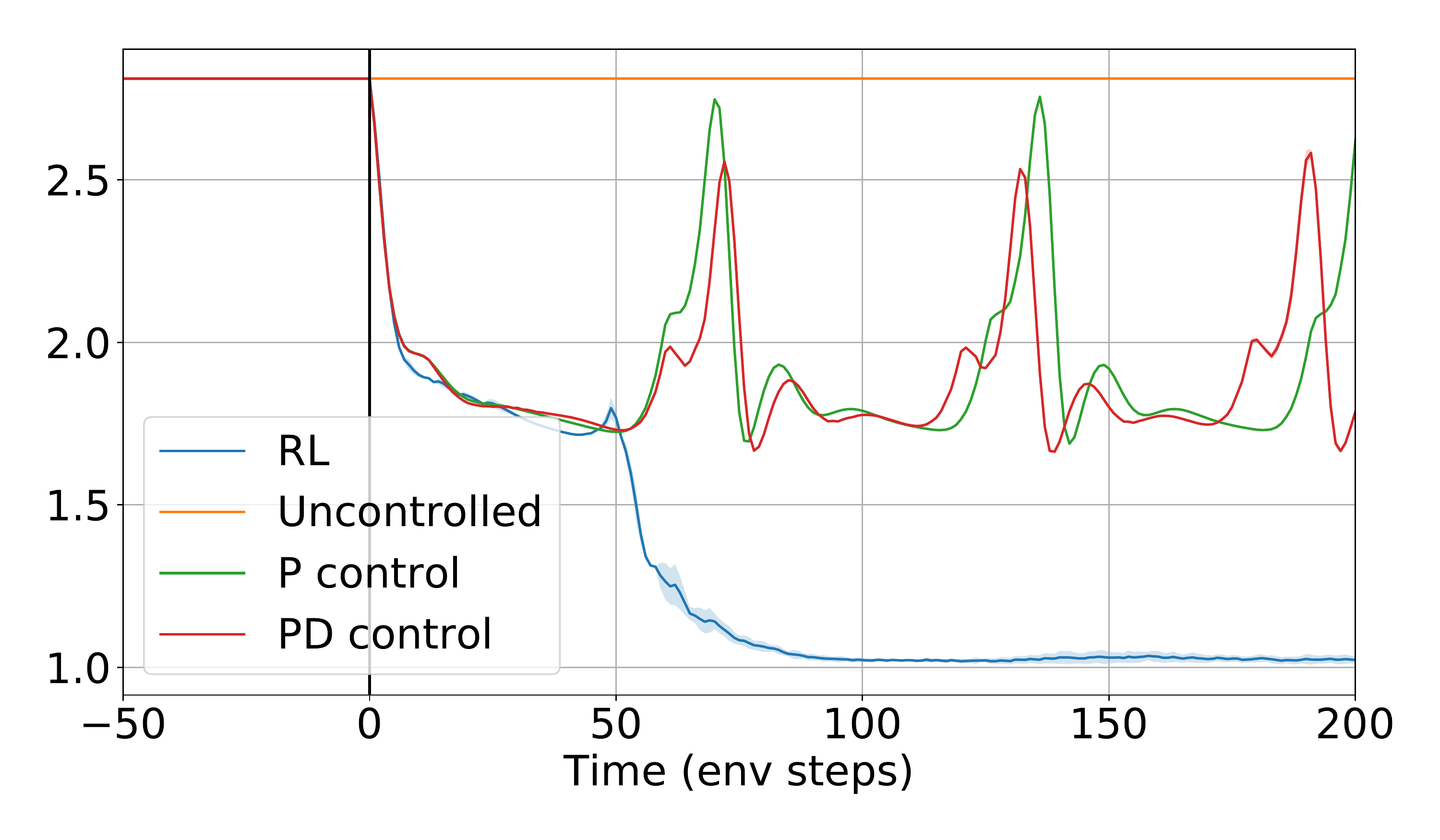}
  \caption{$\text{Ra} = 3 \cdot 10^4$}
  \label{fig:time-3e4}
\end{subfigure}
\begin{subfigure}{.5\textwidth}
  \centering
  \includegraphics[width=1.0\linewidth]{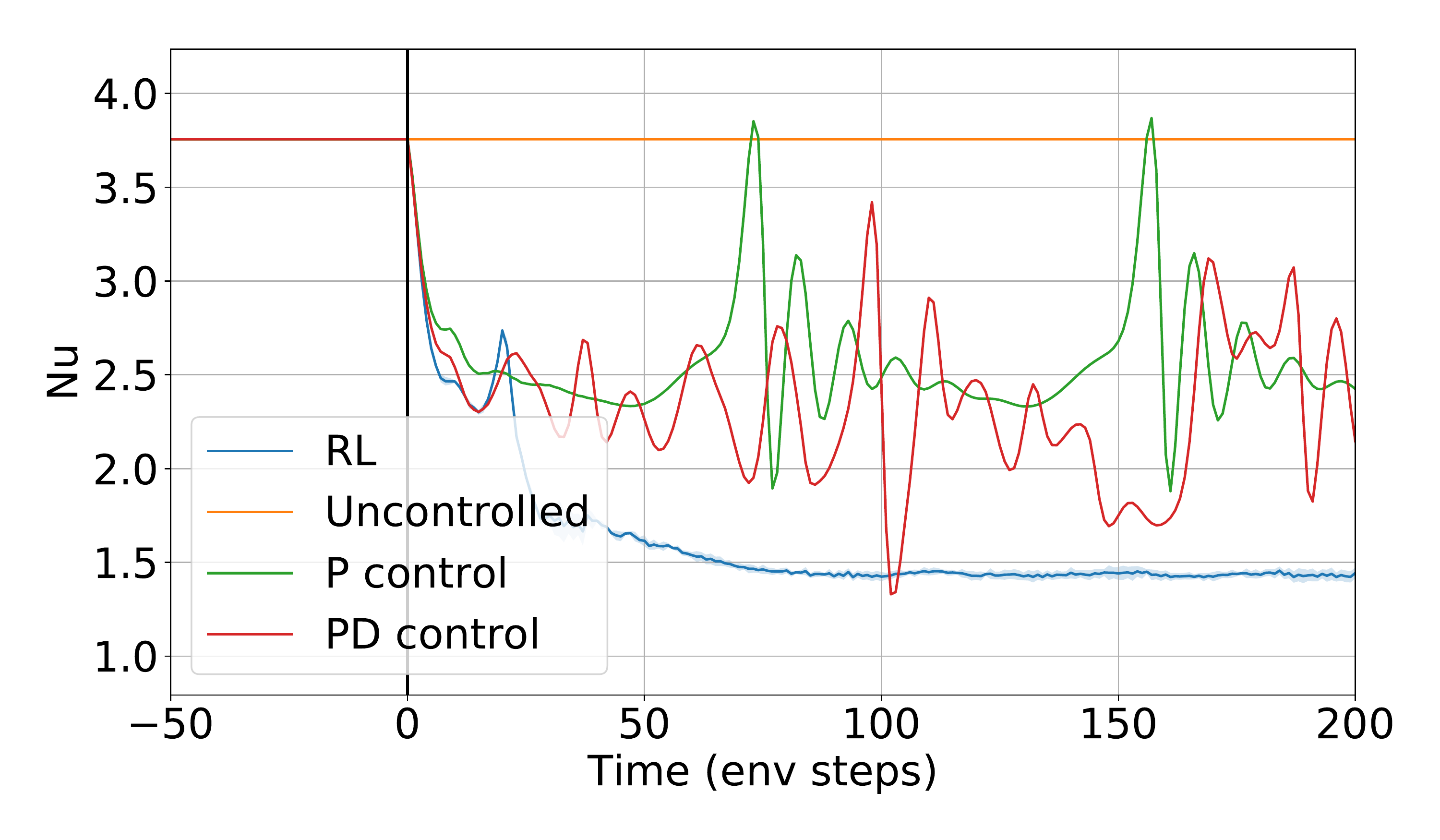}
  \caption{$\text{Ra} = 1 \cdot 10^5$}
  \label{fig:time-1e5}
\end{subfigure}%
\begin{subfigure}{.5\textwidth}
  \centering
  \includegraphics[width=1.0\linewidth]{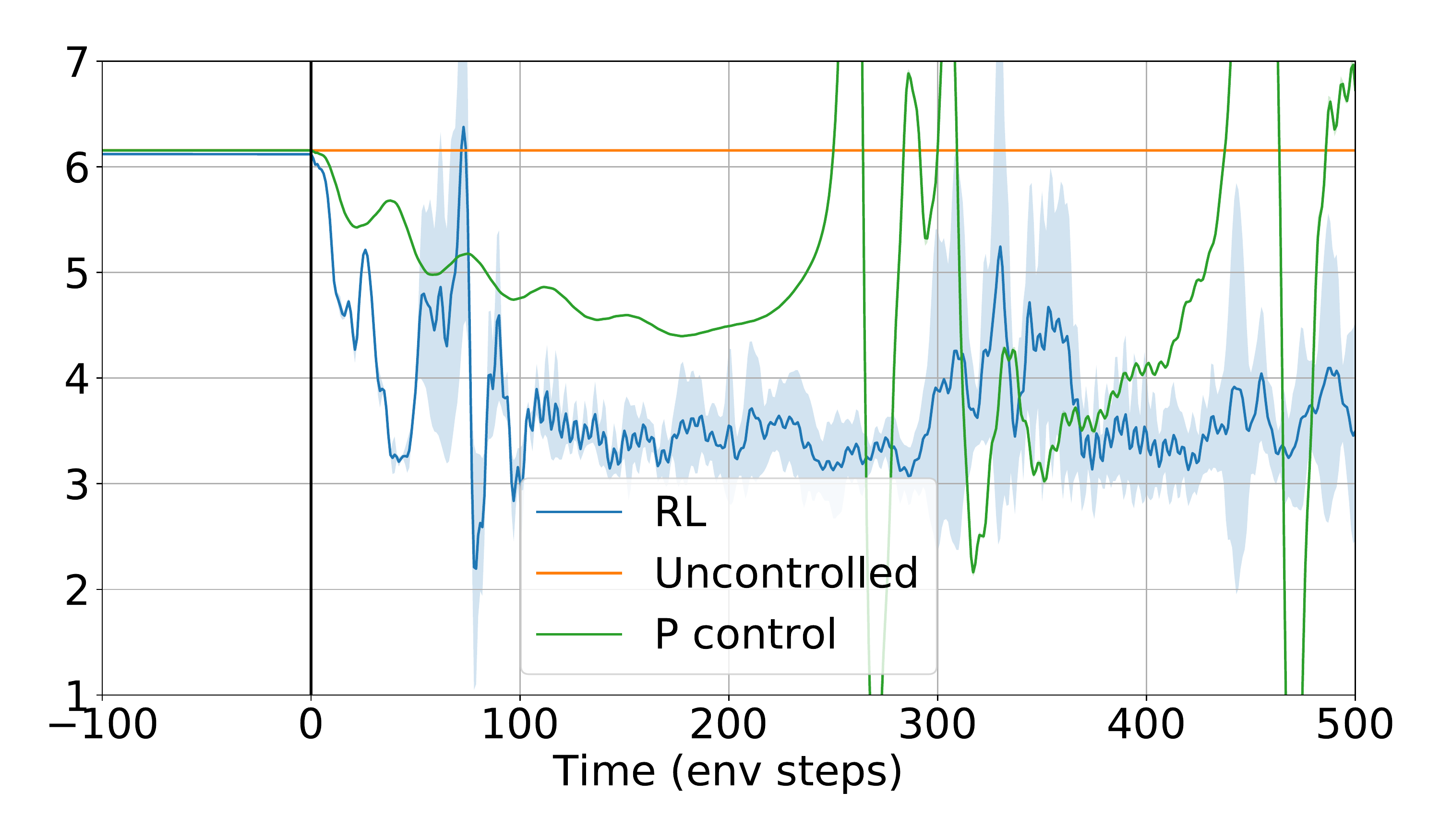}
  \caption{$\text{Ra} = 1 \cdot 10^6$}
  \label{fig:time-1e6}
\end{subfigure}
\caption{Time evolution of the Nusselt number at four different
  Rayleigh regimes, with the control starting at $t=0$. The time axis
  is in units of control loop length, $\Delta t$ (cf. Table
  \ref{tab:RB-envs}). Up to $\text{Ra} = 3 \cdot 10^4$ (a,b), the RL
  control is able to stabilize the system (i.e.
  $\text{Nu} \approx 1$), which is in contrast with linear methods
  that result in a unsteady flow. At $\text{Ra} = 10^5$ (c), the RL
  control is also unable to fully stabilize the system, yet,
  contrarily to the linear case, it still results in a flow having
  stationary \Nu. For $\text{Ra} = 10^6$ (d) the performance of RL is
  not as stable as at lower $\text{Ra}$, the control however still
  manages to reduce the average Nusselt number significantly.}
\label{fig:time-Nu}
\end{figure}

We compare the linear and RB-based control methods on $9$ different
scenarios with Rayleigh number ranging between $\Ra{} = 10^3$ (just
before the onset of convection) and $\Ra{} = 10^7$ (mild turbulence,
see Table~\ref{tab:RB-envs}). Figure~\ref{fig:Ra-Nu-Main} provides a
summary of the performance of the methods in terms of the (ensemble)
averaged Nusselt
number. %
Until $\text{Ra} \approx 10^4$, the RL control and the linear control
deliver similar performance. At higher \Ra{} numbers, in which the \rb
system is more complex and chaotic, RL controls significantly
outperform linear methods. This entails an increment of the critical
Rayleigh number, which increases from $\approx 10^3$, in the
uncontrolled case, to $\approx 10^4$, in case of linear control, and
to $\approx 3\cdot 10^4$ in case of RL-based controls.
Above the critical Rayleigh, RL controls manage a reduction of the
Nusselt number which remains approximately constant, for our specific
experiment, it holds
\begin{equation*}
  \Nu_{\text{uncontrolled}} -\Nu_{\text{RL}} \approx 2.5 \qquad \text{for}\   \Ra{} < 3 \cdot 10^6.
\end{equation*}
In contrast, the linear method scores only
\begin{equation*}
  \Nu_{\text{uncontrolled}} - \Nu_{\text{linear}} \approx 1.5 \qquad \text{for}\   \Ra{} < 10^6,
\end{equation*}
while at higher Rayleigh it results completely ineffective.

\begin{figure}[t]
\centering
\begin{subfigure}{1.0\textwidth}
  \centering
  \includegraphics[width=1.0\linewidth]{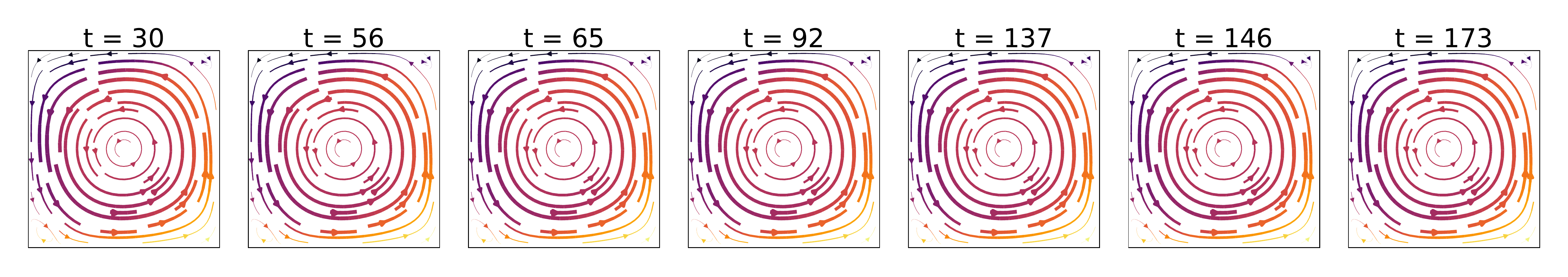}
  \caption{Uncontrolled}
  \label{fig:Flow-Ra1e5-free}
\end{subfigure}
\begin{subfigure}{1.0\textwidth}
  \centering
  \includegraphics[width=1.0\linewidth]{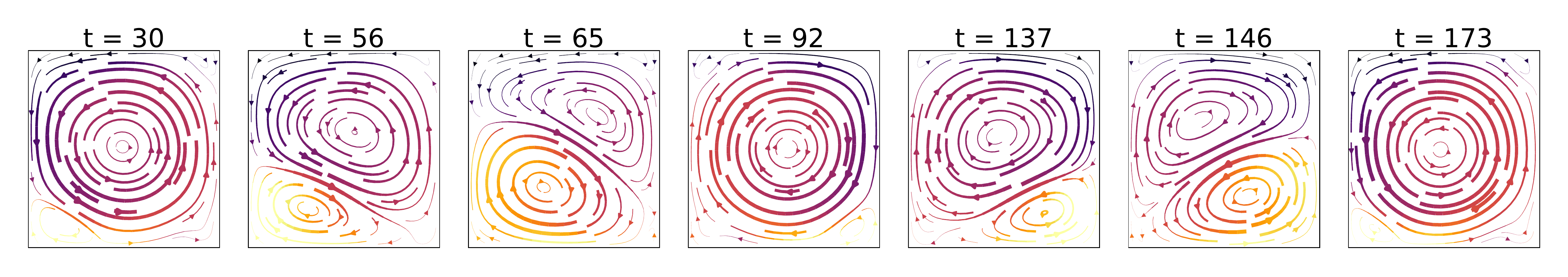}
  \caption{Linear proportional controlled}
  \label{fig:Flow-Ra1e5-G}
\end{subfigure}
\begin{subfigure}{1.0\textwidth}
  \centering
  \includegraphics[width=1.0\linewidth]{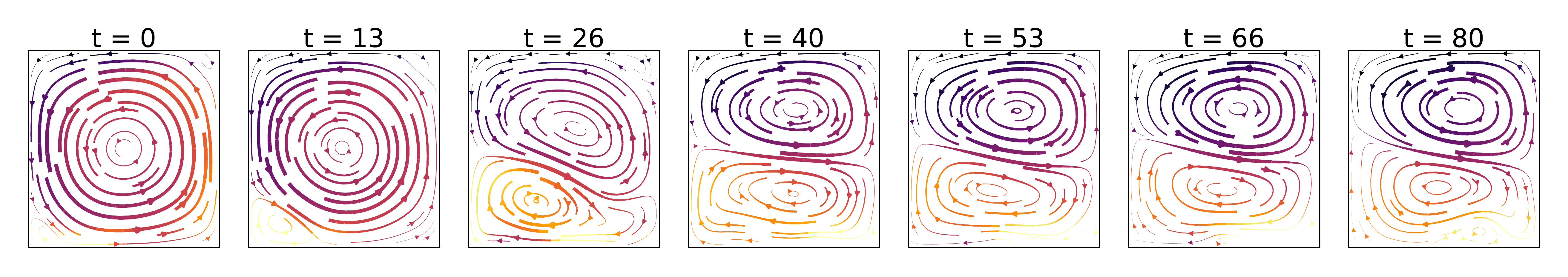}
  \caption{RL controlled}
  \label{fig:Flow-Ra1e5-RL}
\end{subfigure}
\caption{Instantaneous stream lines at different times, at $\text{Ra} =  10^5$, comparison of cases  without control (a),  with linear control (b), and with RL-based control (c). Note that the time is given in units of $\Delta t$ (i.e control loop length, cf. Table~\ref{tab:RB-envs}. Note that the snapshots are taken at different instants).
  RL controls establishes a flow regime like a ``double cell mode''
  which features a steady Nusselt number (see
  Figure~\ref{fig:time-1e5}). This is in contrast with linear methods
  which rather produce a flow with fluctuating Nusselt. The double
  cell flow field has a significantly lower Nusselt number than the
  uncontrolled case, as heat transport to the top boundary can only
  happen via diffusion through the interface between the two
  cells. This ``double cell'' control strategy is established by the
  RL control with any external supervision.}
\label{fig:Flow-Ra1e5}
\end{figure}

\begin{figure}[t]
\centering
\begin{subfigure}{1.0\textwidth}
  \centering
  \includegraphics[width=1.0\linewidth]{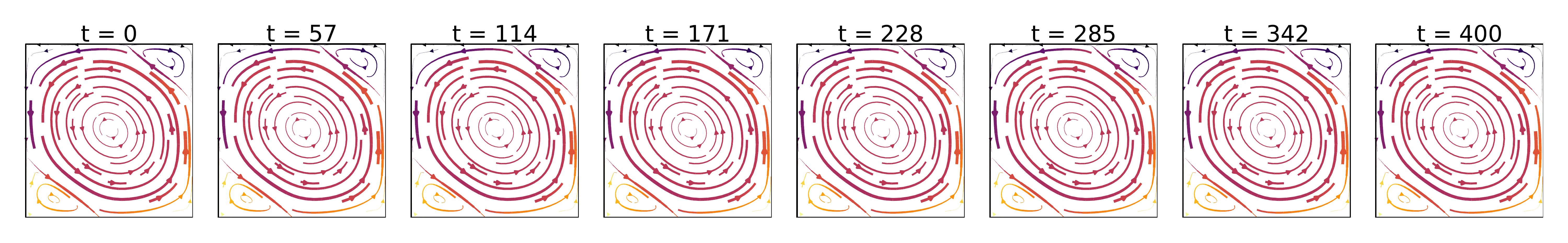}
  \caption{Uncontrolled}
  \label{fig:Flow-Ra3e6-free}
\end{subfigure}
\begin{subfigure}{1.0\textwidth}
  \centering
  \includegraphics[width=1.0\linewidth]{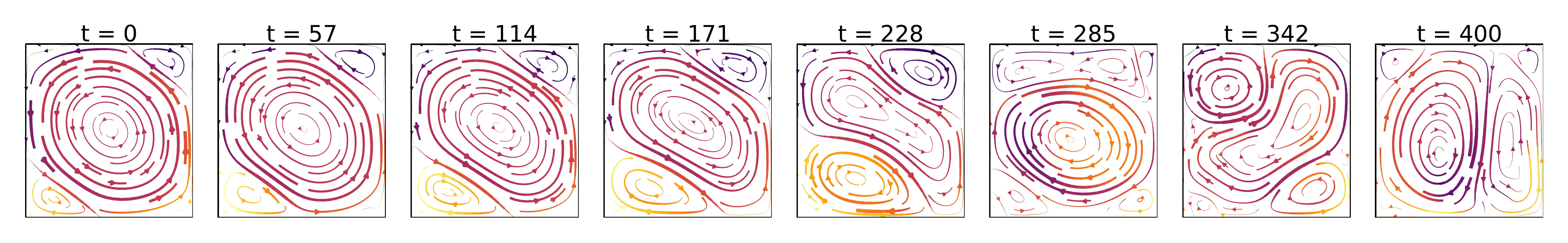}
  \caption{Linear proportional controlled}
  \label{fig:Flow-Ra3e6-G}
\end{subfigure}
\begin{subfigure}{1.0\textwidth}
  \centering
  \includegraphics[width=1.0\linewidth]{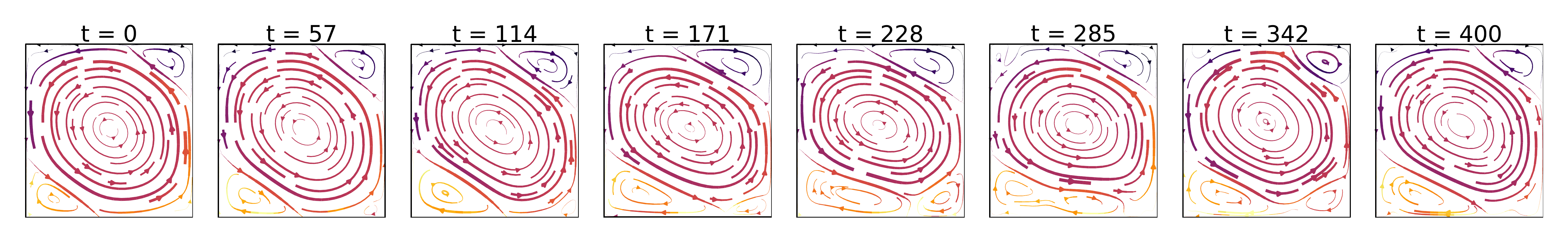}
  \caption{RL controlled}
  \label{fig:Flow-Ra3e6-RL}
\end{subfigure}
\caption{Instantaneous stream lines at
  $\text{Ra} = 3 \cdot 10^6$, comparison of cases without control (a),
  with linear control (b), and with RL-based control (c). 
  We observe that the RL control still tries to enforce a ``double
  cell'' flow, as in the lower Rayleigh case (Figure~\ref{fig:Flow-Ra1e5}).
  The structure is however far less pronounced and convective effects
  are much stronger. This is likely due to the increased instability
  and chaoticity of the system, which increases the learning and
  controlling difficulty. Yet, we can observe a small alteration in
  the flow structure (cf. lower cell, larger in size than in
  uncontrolled conditions) which results in a slightly lower Nusselt number.}
\label{fig:Flow-Ra3e6}
\end{figure}

The reduction of the average Nusselt number is an indicator of the
overall suppression -or the reduction- of convective effects, yet it
does not provide insightful and quantitative information on the
characteristics of the control and of the flow. In
Figure~\ref{fig:time-Nu}, we compare the controls in terms of the time
histories of the (ensemble-averaged) Nusselt number. We include four
different scenarios. For $\text{Ra} = 3 \cdot 10^4$ the RL control is
able to stabilize the system ($\text{Nu} \approx 1$) while both linear
control methods result in periodic orbits~\cite{intro:PDRBcontrol}. At
$\text{Ra} = 10^5$, RL controls are also unable to stabilize the
system; yet, this does not result in any periodic flows as in the case
of linear control. Finally, at $\text{Ra} = 10^6$ we observe a
time-varying Nusselt number even using RL-based controls. To better
understand the RL control strategy, in Figures~\ref{fig:Flow-Ra1e5}
and~\ref{fig:Flow-Ra3e6}, we plot the instantaneous velocity
streamlines for the last two scenarios.

For the case $\text{Ra} = 10^5$ (see Figure~\ref{fig:Flow-Ra1e5}(c)),
the RL control steers and stabilizes the system towards a
configuration that resembles a double cell. This reduces convective
effect by effectively halving the effective Rayleigh number. In
particular, we can compute a effective Rayleigh number,
$\Ra{}_{\text{eff}}$, by observing that the double cell structure can
be constructed as two vertically stacked \rb systems with halved
temperature difference and height (i.e., in formulas,
$\Delta T' = \Delta T/2$ and $H' = H/2$). It thus results in an
effective Rayleigh number satisfying
 \begin{equation}
   \Ra{}_{\text{eff}}
  = \frac{g \alpha \Delta T' H'^3}{\nu \kappa} =
   \frac{1}{16}\frac{g \alpha \Delta T H^3}{\nu \kappa} = \frac{1}{16}
   \Ra{},
 \end{equation}
which is in line with the observed reduction in Nusselt.

At $\text{Ra} = 3 \cdot 10^6$ it appears that the RL control attempts,
yet fails, to establish the ``double cell'' configuration observed
at lower $\Ra{}$ (cf. Figure~\ref{fig:Flow-Ra3e6}(c)).  Likely, this is
connected to the increased instability of the double cell
configuration as Rayleigh increases.

These results were achieved with less than an hour of training time on
an Nvidia V100 for the cases with low Rayleigh number
($\text{Ra} \lesssim 10^5$). However, at $\text{Ra} \gtrsim 10^6$ the
optimization took up to 150 hours of computing time (the majority of
which is spent in simulations and the minority of which is spent in
updating the policy). For further details on the training process of
the RL controller see Appendix~\ref{sec:appendix-training}.

\section{Limits to learning and control}\label{sec:limits}
In this section we discuss possible limits to the capability of
learning automatic controls when aiming at suppressing convection. Our
arguments are grounded on general physics
concepts and thus apply to other control problems 
for non-linear/chaotic dynamics.

In Sect.~\ref{sec:rb-control-background-and-implementation}, 
we showed that RL controls are capable
of substantially outperforming linear methods in presence
of sufficient control complexity ($\Ra\gtrsim 10^4$). 
It remains however unclear 
how far these controls are from optimality, especially at high Ra. 
Here we address the physics factors certainly
hindering learning and controlling capabilities.

Having sufficient time and spatial resolution on the relevant state of the system 
is an obvious requirement to allow a successful control. Such resolution
however is not defined in absolute terms, rather it connects to the
typical scales of the system and, in case of a chaotic behavior,
with the Lyapunov time and associated length scale. In our case, at fixed \Ra{},
learnability and controllability connect with the number and density of probes, 
with the time and space resolution of the control, but also with its
``distance'' with respect to the bulk of the flow.

The system state is observed via a number of probes in fixed position
(see Figure~\ref{fig:RL-scheme}). For a typical (buoyant) velocity in
the cell, $v_b$, there is a timescale associated with the delay with
which a probe records a sudden change (e.g. creation/change of a
plume) in the system. When this timescale becomes comparable or larger
than the Lyapunov time, it appears hopeless for any control to learn
and disentangle features from the probe readings. In other words, as
$\Ra$ increases, we expect that a higher and higher number of probes
becomes necessary (but not sufficient) in order to learn control
policies.

Besides, our choice to control the system via the bottom plate
temperature fluctuations entails another typical delay: the time taken
to thermal adjustments to propagate inside the cell.  In particular,
in a cell at rest, this is the diffusion time,
$t_D \sim \frac{H^2}{\kappa} \sim \Ra{}^{1/2}$.  If the delay gets
larger or comparable to the Lyapunov time, controlling the system
becomes, again, likely impossible.

\begin{figure}[h]
\setlength{\unitlength}{10cm}
\begin{center}
\includegraphics[width=10cm]{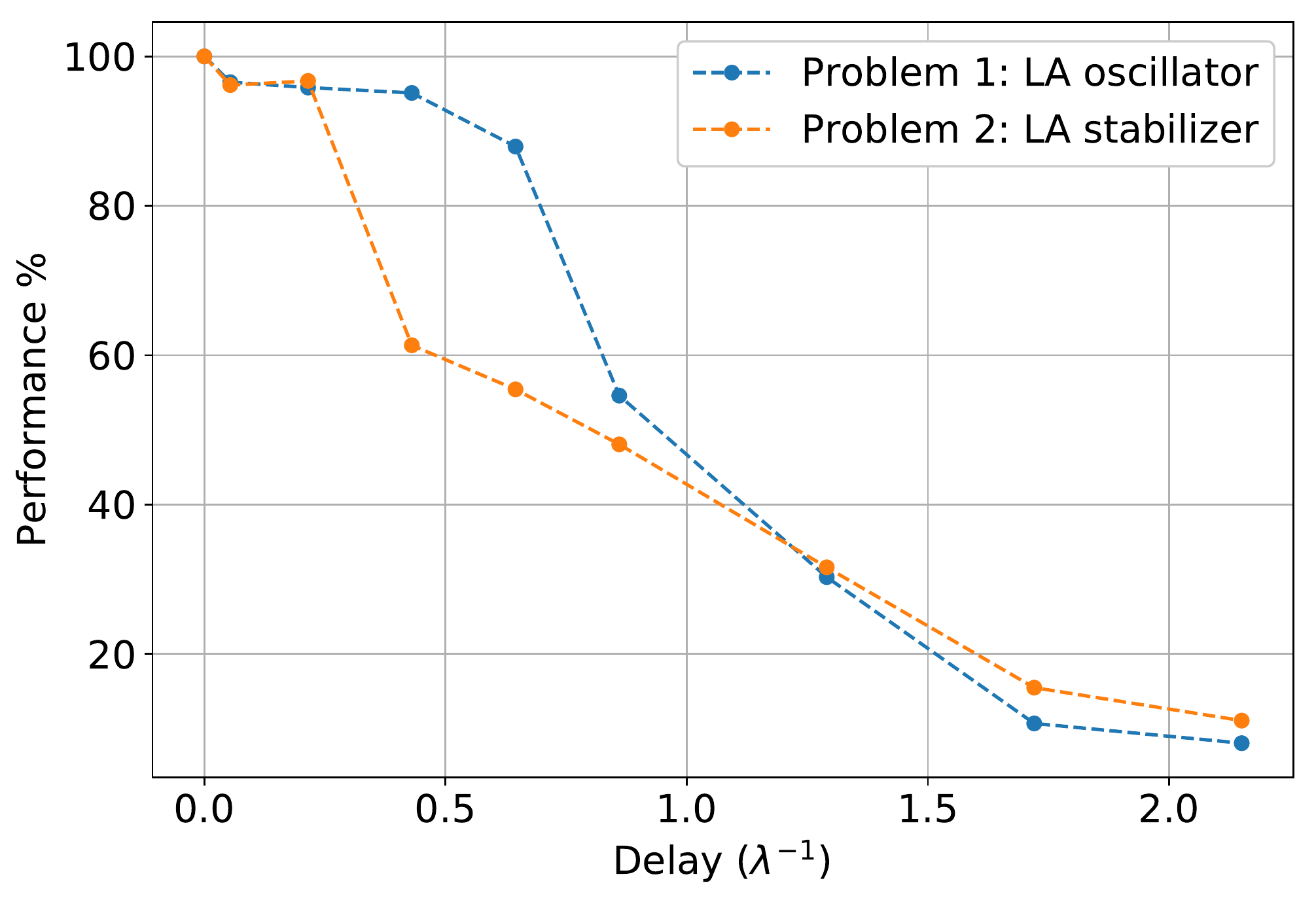}
\end{center}
\caption{Performance loss due to an artificial delay imposed to a RL
  controller operating on the Lorenz attractor. The controller,
  operating on the $y$ variable of the system (reduced model for the
  RBC horizontal temperature fluctuations) aims at either maximizing
  (LA oscillator) or minimizing (LA stabilizer) the number of sign
  changes of the $x$ (in RBC terms the angular velocity of the flow).
  The delay, on the horizontal axis, is scaled on the Lyapunov time,
  $\lambda^{-1}$, of the system (with $\lambda$ the largest Lyapunov
  exponent).  In case of a delay in the control loop comparable in
  size to the Lyapunov time, the control performance diminishes
  significantly. }
\label{fig:LA-delay-performance}
\end{figure}
To illustrate this concept, we rely on a well-known low-dimensional
model inspired by RBC: the Lorenz attractor. The attractor state is
three-dimensional and its variables (usually denoted by $x,y,z$; see
Appendix~\ref{sec:LA}) represent the flow angular velocity, the
horizontal temperature fluctuations and the vertical temperature
fluctuations.  We consider a RL control acting on the horizontal
temperature fluctuations ($y$ variable) that aims at either minimizing
or maximizing the sign changes of the angular velocity
(i.e. maximizing or minimizing the frequency of sign changes of the
$x$ variable). In other words, the control aims at keeping the flow
rotation direction maximally consistent or, conversely, at magnifying
the rate of velocity inversions.  In this simplified context, we can
easily quantify the effects of an artificial delay in the control on
the overall control performance
(Figure~\ref{fig:LA-delay-performance}).  Consistently with our
previous observations, when the artificial delay approaches the
Lyapunov time the control performance significantly degrades.

\begin{figure}[h]
\setlength{\unitlength}{10cm}
\begin{center}
\includegraphics[width=10cm]{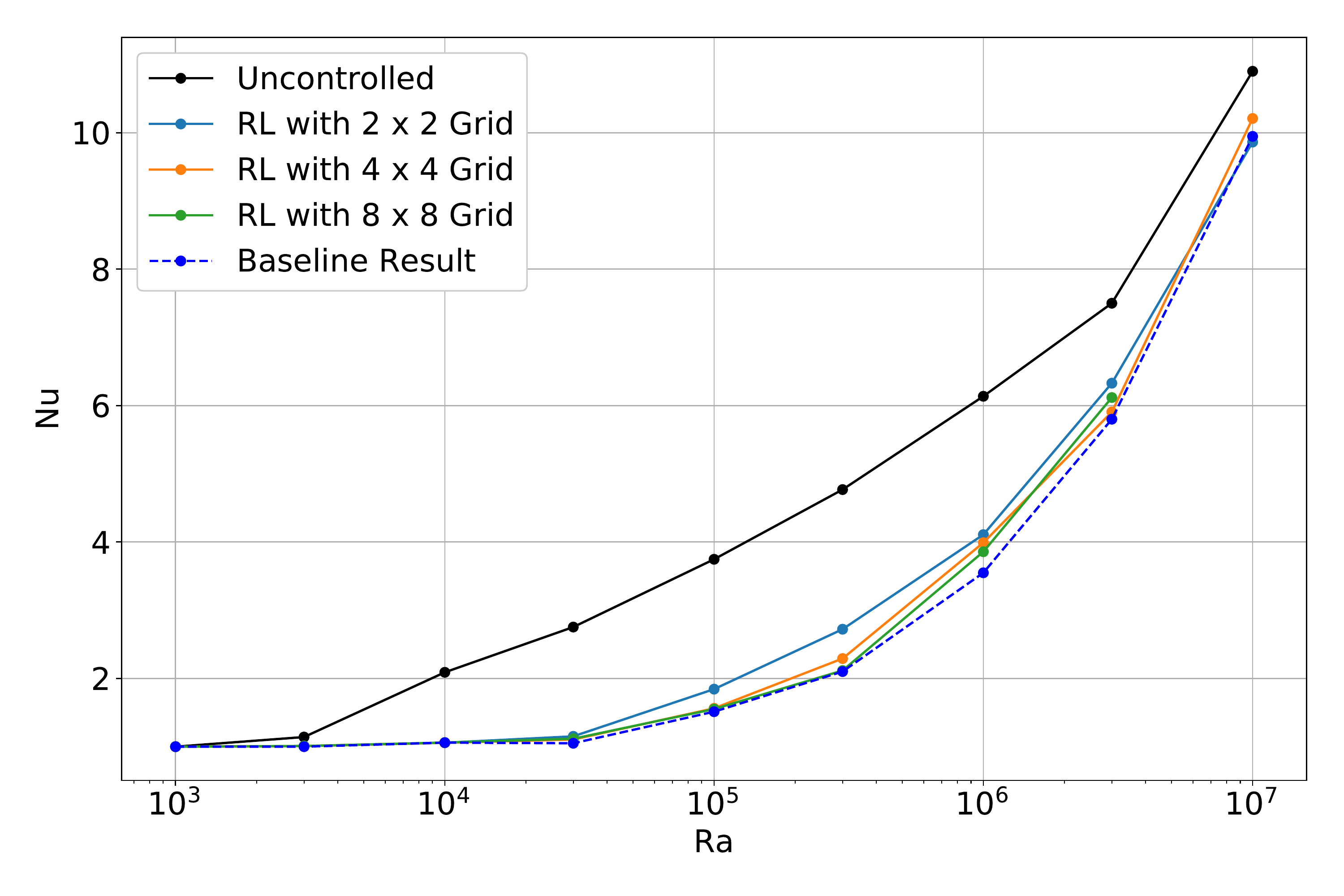}
\end{center}
\caption{Average Nusselt number for a RL agent observing the \rb
  environment at different probe density, all of which are lower than
  the baseline employed in Section~\ref{fig:Ra-Nu-Main}. The grid
  sizes (i.e information) used to sample the state of the system at
  the \Ra{} considered does not seem to play key role in limiting the
  final control performance of the RL agent.}
\label{fig:Probed-Ra-Nu}
\end{figure}

Notably, in our original 2D RBC control problem
(Sect.~\ref{sec:rb-control-background-and-implementation}), control
limitations are not connected to delays in observability. In fact, as
we consider coarser and coarser grids of probes, the control
performance does not diminish significantly
(cf. Fig.~\ref{fig:Probed-Ra-Nu}; surprisingly, observations via only
$4$ allow similar control performance to what achieved employing $64$
probes).  This suggests that other mechanisms than insufficient
probing determine the performance, most likely, the excessively long
propagation time (in relation to the Lyapunov time) needed by the
controlling actions to traverse the cell from the boundary to the
bulk. This could be possibly lowered by considering different control
and actuation strategies.

\section{Discussion} \label{sec:conclusion}

In this paper we considered the issue of suppressing convective
effects in a 2D \rb{} cell, by applying small temperature fluctuations
at the bottom plate. In our proof of concept comparison, limited to a
square cell and fixed $\Pran$, we showed that controls based on
Reinforcement Learning (RL) are able to significantly outperform
state-of-the-art linear approaches.  Specifically, RL is capable of
discovering controls stabilizing the \rb system up to a critical
Rayleigh number that is approximately $3$ times larger than achievable
by linear controls and $30$ times larger than in the uncontrolled
case.  Secondly, when the purely conductive state could not be
achieved, the RL still produces control strategies capable of reducing
convection, which are significantly better than linear algorithms. The
RL control achieves this by inducing an unstable flow mode, similar to
a stacked double-cell, yet not utilized in the context of RBC control.

Actually no guarantee exists on the optimality of the controls found
by RL. Similarly it holds for the linear controls, which additionally
need vast manual intervention for the identification of the
parameters.  However, as we showed numerically, theoretical bounds to
controllability hold which are regulated by the chaotic nature of the
system, i.e. by its Lyapunov exponents, in connection with the (space
and time) resolution of the system observations as well as with the
actuation capabilities. We quantified such theoretical bounds in terms
of delays, in observability and/or in actuation: whenever these become
comparable to the Lyapunov time, the control becomes impossible.

There is potential for replication of this work in an
actual experimental setting. However, training a controller only via experiments might take an excessively long time to converge. Recent
developments in reinforcement learning showed already the possibility of
employing controllers partially trained on simulations (transfer
learning~\cite{intro:RL-example-irl-cube}).
This would not only be a large step for the control of flows, but also
for RL where practical/industrial uses are still mostly
lacking~\cite{intro:RL-book}.

\appendix

\section{\rb simulation details} \label{sec:appendix-RB-sim}

\note{Simulation specs} We simulate the \rb system via the lattice
Boltzmann method (LBM) that we implement in a vectorized way on a
GPU. We employ the following methods and schemes
\begin{itemize}
\item \textit{Velocity population:} D2Q9 scheme, afterwards indicated by $f_i(\mathbf{x},t)$;
\item \textit{Temperature population:} D2Q4 scheme;%
\item \textit{Collision model:} BGK collision operator~\cite{Appendix:BGK,intro:lbm-book};
\item \textit{Forcing scheme:} As seen in~\cite{intro:lbm-book} most forcing schemes can be formulated as follows
\begin{align}
& f_i(\mathbf{x} + \mathbf{e}_i \Delta t,t + \Delta t) - f_i(\mathbf{x},t) = \left[\Omega_i(\mathbf{x},t) + S_i(\mathbf{x},t) \right] \Delta t \\
& \mathbf{u}^{\text{eq}} = \frac{1}{\rho} \sum_i f_i \mathbf{e}_i + A \frac{\mathbf{F} \Delta t}{\rho}
\end{align}
with $S_i$ and $A$ defined by scheme. We choose the scheme by He et al.~\cite{Appendix:He-force-model} for its improved accuracy which reads
\begin{align}
A &= \frac{1}{2} \\
S_i &= \left(1-\frac{\Delta t}{2 \tau}\right) \frac{f_i^{\text{eq}}}{\rho} \frac{\mathbf{e}_i-\mathbf{u}}{c_s^2} \mathbf{F}
\end{align}
\item \textit{Boundary model:} bounce-back rule enforcing no-slip boundary conditions~\cite{intro:lbm-book}.
\end{itemize}
To limit the training time, we implemented the LBM vectorizing on the
simulations. This enabled us to simulate multiple concurrent, fully
independent, \rb systems within a single process. This eliminates the
overhead of having numerous individual processes running on a single
GPU which would increase the CPU communication overhead.

\note{1 env step and episode setup}

When a RL controller selects a temperature profile for the bottom
boundary this is endured for number of LBM steps (this defines one
environment step, or env step, with length $\Delta t$).  The reason
for these, so-called, sticky actions is that within one env step the
system does not change significantly. Allowing quicker actions would
not only be physically meaningless but also possibly detrimental to
the performance (this is a known issue when training RL agents for
games where actions need to be sustained to make an
impact~\cite{intro:RL-example-Curiosity-large}).
Furthermore, due to our need to investigate the transient behavior, we
set the episode length to $500$ env steps. In this way, the transient
is extinguished within the first $150$ env steps. After each episode
the system is reset to a random, fully developed, convective RB state.

In dependence on the Rayleigh number (i.e. system size), it takes between millions and billions env steps to converge to a control strategy. To limit the computing time, we consider the smallest possible system that gives a good estimate for the uncontrolled
Nusselt number (error within few percent).

\note{table results}

In Table~\ref{tab:RB-envs} we report the considered Rayleigh numbers and related system sizes.

\begin{table}[H]
\centering
\caption{\rb environments considered. For each Rayleigh number we
  report the LBM grid employed (size $N_X\times N_Y$), the
  uncontrolled Nusselt number measured from LBM simulations and a
  validation reference from the
  literature.} %
\begin{tabular}{l|llll}
\hline
  \Ra{}             & $N_X$ and $N_Y$   & \begin{tabular}[x]{@{}c@{}}Length 1 env step \\ (i.e. control loop length)\\ (units: lbm steps) \end{tabular}   & \Nu     & \Nu{} reference  \cite{Appendix:RB-Nu-ref}  \\
\hline
 $1 \cdot 10^3$ & 20          & 16                              & 1.000  & 1.000  \\
 $3 \cdot 10^3$ & 20          & 16                              & 1.141  &        \\
 $1 \cdot 10^4$ & 20          & 30                              & 2.090  & 2.15   \\
 $3 \cdot 10^4$ & 25          & 60                              & 2.753  &        \\
 $1 \cdot 10^5$ & 30          & 60                              & 3.847  & 3.91  \\
 $3 \cdot 10^5$ & 40          & 60                              & 4.768  &        \\
 $1 \cdot 10^6$ & 100         & 60                              & 6.136  & 6.3    \\
 $3 \cdot 10^6$ & 200         & 100                             & 7.500  &        \\
 $1 \cdot 10^7$ & 350         & 180                             & 10.900 &        \\
\hline
\end{tabular}
\label{tab:RB-envs}
\end{table}

\section{Control amplitude normalization}\label{sec:app-normaliz}
To limit the amplitude of the temperature fluctuations and ensure
their zero-average (see
Eq.~\eqref{eq:control-constraint-average},~\eqref{eq:bound-temp-fluct})
we employ the following three normalization steps, indicated by
$\mathcal{R}(\cdot)$ in the manuscript. Let $\tilde{T}_B(x,t)$ be the
temperature fluctuation proposed by either the linear control or the
RL-based control, we obtain $\hat{T}_B(x,t)$ as
\begin{align}\label{eq:normalization-temp}
    & \tilde T_{B}'(x,t) = \mbox{Clip}(\tilde T_{B}(x,t),-C,C) \\
    & \tilde T_{B}''(x,t) = \tilde T_{B}'(x,t) - \langle \tilde T_{B}'(x,t) \rangle_x \\
    & \hat T_B(x,t) = \frac{\tilde T_{B}''(x,t)}{\max_{x'}(1,|\tilde T_{B}''(x,t)|/C)}.
\end{align}
Note that the first operation is a clipping of the local temperature
fluctuation between $\pm C$, which is necessary only for the linear
control case.

\section{RL algorithm implementation and hyperparameters} \label{sec:appendix-RL-implement}

In this appendix we elaborate on our design choices about the implementation of RL for \rb control.

\begin{itemize}
\item \textit{Discretization of the bottom boundary in 10 sections.}
  A literature
  study~\cite{LAres:RL-continues-action-2,LAres:RL-continues-action}
  and preliminary experiments have shown that large/continuous action
  spaces are currently rather challenging for the convergence of RL
  methods. In our preliminary experiments we observed that
  discretizing $T_B$ in $20$ sections was even less effective that in
  $10$ sections, and that $5$ sections were instead insufficient to
  get the desired performance.
\item \textit{3 layer multilayer perceptron (MLP) for state encoding.}
  We considered this option over a convolutional neural network (CNN)
  applied on the entire lattice. The latter had significantly longer
  training times and lower final performance. Besides, we included in
  the state observed by the MLP the system readings in the $D$
  previous env steps, which is known to be beneficial for
  performance~\cite{intro:RL-example-Curiosity-large}.
\item \textit{PPO algorithm.} We considered this option over
  value-based methods which were however more difficult to operate
  with due to the need of extensive hyperparameter
  tuning. Furthermore, we used the open source implementation of PPO
  included in the stable-baselines python
  library~\cite{intro:RL-platform-stable-baselines} (note: training
  PPO demands for a so-called auxiliary value
  function~\cite{intro:PPO-algorithm}. For that we employed a separate
  neural network having the same structure as the policy function).
\end{itemize}

\subsection{Hyperparameters}

We used the work by Burda et
al.~\cite{intro:RL-example-Curiosity-large} as a starting point for
our choice of the hyperparameters. We specifically considered two
separate hyperparameter sets. \textbf{i} targeting final performance
over training speed, used for $\text{Ra} \leq
10^6$. \textbf{ii} targeting speed over final performance, 
used only on the highest Rayleigh number case ($\text{Ra} > 10^6$) and
for the research on the probe density. Below one can see the PPO
hyperparameters used (see~\cite{intro:RL-book}
and~\cite{intro:PPO-algorithm} for further explanations).

\begin{itemize}
\item Number of concurrent environments: 512 (set 2: 128)
\item Number of roll-out steps: 128
\item Number of samples training samples: $512 \cdot 128 = 65536$ (set 2: $128 \cdot 128 = 16384$)
\item Entropy coefficient $c_s$: 0.01
\item Learning rate $\alpha$: $2.5 \cdot 10^{-4}$
\item Discount factor $\gamma$: 0.99
\item Number of mini-batches: 8 (set 2: 16)
\item Number of epoch when optimizing the surrogate: 4 (set 2: 32)
\item Value function coefficient for the loss calculation: 0.5
\item Factor for trade-off of bias vs. variance for Generalized Advantage Estimator $\Lambda$: 0.95
\item PPO Clip-range: 0.2
\end{itemize}

\section{Training curves} \label{sec:appendix-training} 
\begin{figure}[t]
\centering
\begin{subfigure}{.33\textwidth}
  \centering
  \includegraphics[width=.99\linewidth]{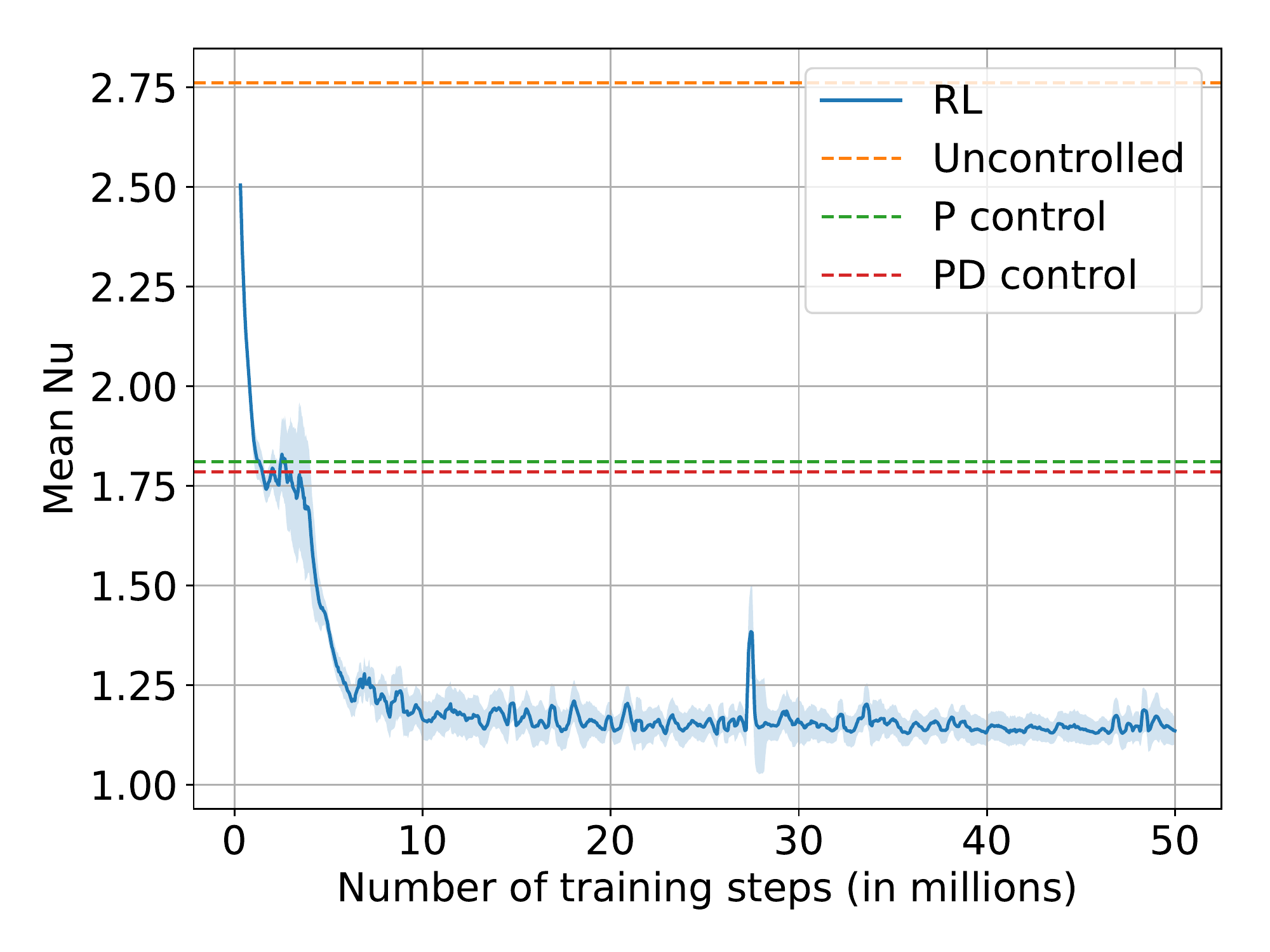}
  \caption{``Low'' \Ra{} of $3 \cdot 10^4$}
  \label{fig:Learning-Nu-Ra-small}
\end{subfigure}%
\begin{subfigure}{.33\textwidth}
  \centering
  \includegraphics[width=.99\linewidth]{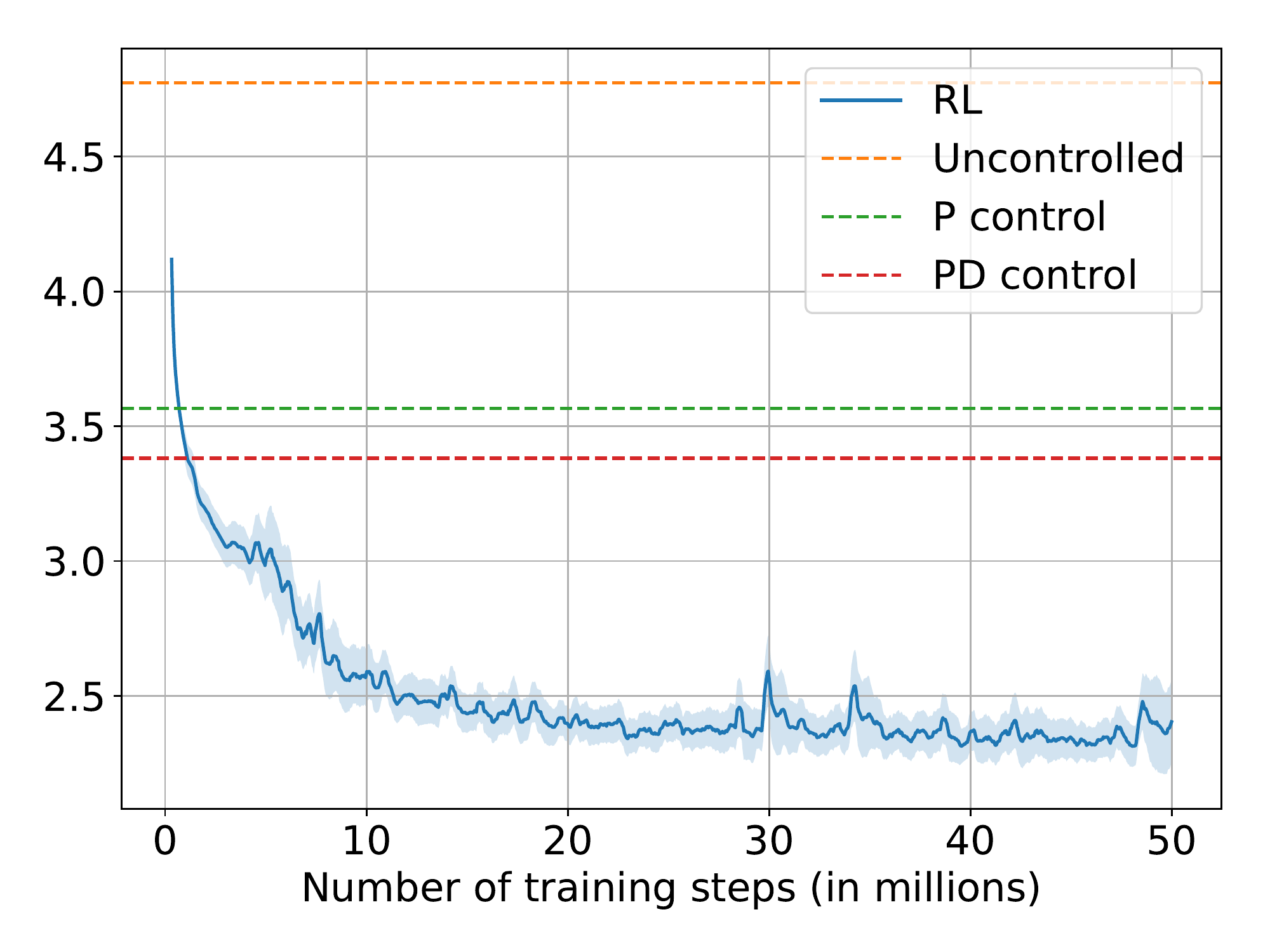}
  \caption{``Mid'' \Ra{} of $3 \cdot 10^5$}
  \label{fig:Learning-Nu-Ra-mid}
\end{subfigure}
\begin{subfigure}{.33\textwidth}
  \centering
  \includegraphics[width=.99\linewidth]{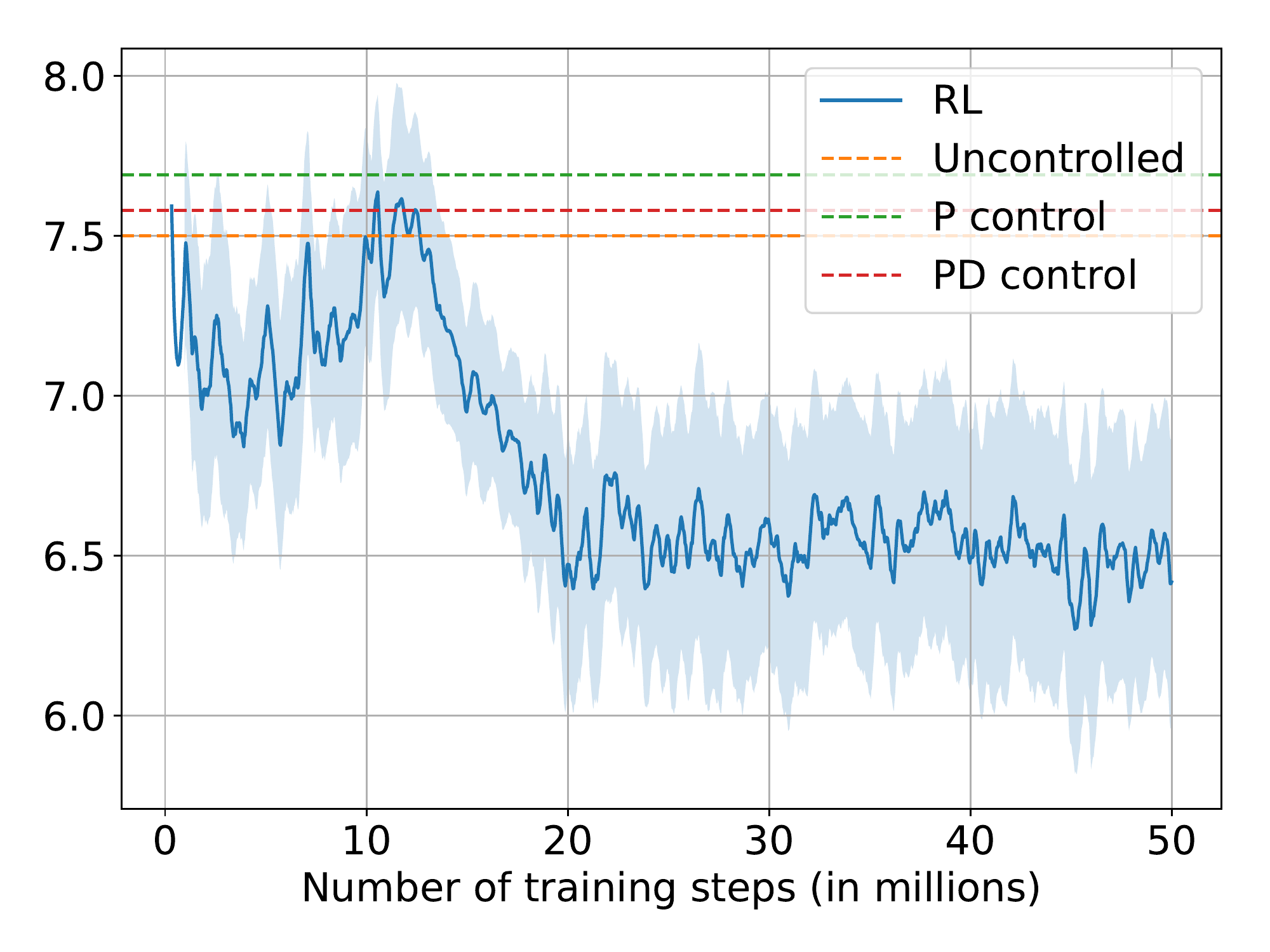}
  \caption{``High'' \Ra{} of $3 \cdot 10^6$}
  \label{fig:Learning-Nu-Ra-high}
\end{subfigure}
\caption{Performance of RL during training (case of the \rb
  system). We report the average Nusselt number and its fluctuations
  computed among a batch of 512 concurrent training environments. (a)
  ``low'' \Ra{} ($0 \lesssim \Ra{} \leq 1 \cdot 5 \lesssim 10^4$) in
  which the control achieves $\text{Nu} \approx 1$ in a stable way,
  (b)``mid'' \Ra{}
  ($5 \cdot 10^4 \lesssim \text{Ra} \leq 1 \lesssim 10^6$) which still
  gives stable learning behavior but converges to $\text{Nu} > 1$ and,
  lastly, (c) ``high'' \Ra{} ($1 \cdot 10^6 \lesssim \Ra{}$) in which
  the higher chaoticity of the system makes a full stabilization
  impossible.}
\label{fig:Learning-Nu-Ra}
\end{figure}

We report in
Figure~\ref{fig:Learning-Nu-Ra} the learning curves for our RL control
(performance vs. length of the training session).
These curves provide information on the time necessary to converge to
a strategy and thus are an indication of the  difficulty and stability of the process. 

\section{Implementation Lorentz Attractor Control}
\label{sec:LA}

To illustrate our argument that a delay comparable to the Lyapunov
time is detrimental to the control performance, we introduce two
control problems defined on the Lorentz Attractor (LA). These LA
control problems are defined considering the following equations
 \begin{align}
 & \dot x = \sigma (y - x), \\
 & \dot y = x (\rho - z) - y + a, \\
 & \dot z = x y - \beta z, \\
 & \text{subject to} \ \  |a| \leq 1
 \label{eq:LA-controlled}
 \end{align}
 with $a$ being a relatively small controllable parameter, and
 $\sigma = 10$, $\rho = 28$ and $\beta = 8/3$. The control loop and
 integration loop (via RK4) have the same time stepping
 $\Delta t = 0.05$.  The two control problems are as follows
 \begin{enumerate}
 \item \textit{``LA stabilizer''}. We aim at minimizing the frequency
   with which the flow direction changes (i.e. the frequency of $x$
   sign changes). Reward: $R_i = - 1$ if $x_{i-1} x_i < 0$ and zero otherwise;
     \item \textit{``LA oscillator''}. Similar to LA stabilizer but
       with inverse goal. Reward: $R_i = + 1$ if $x_{i-1} x_i < 0$ and
       zero otherwise.
 \end{enumerate}
  We start the system in a random state around the attractor, the controller is an MLP network, and we use the PPO RL algorithm (similarly to our approach for the complete \rb case). We furthermore limit the control to three states, $a = -1 \vee 0 \vee 1$, for training speed purposes. 

\begin{figure}[t]
\centering
\begin{subfigure}{1.0\textwidth}
  \centering
  \includegraphics[width=0.66\linewidth]{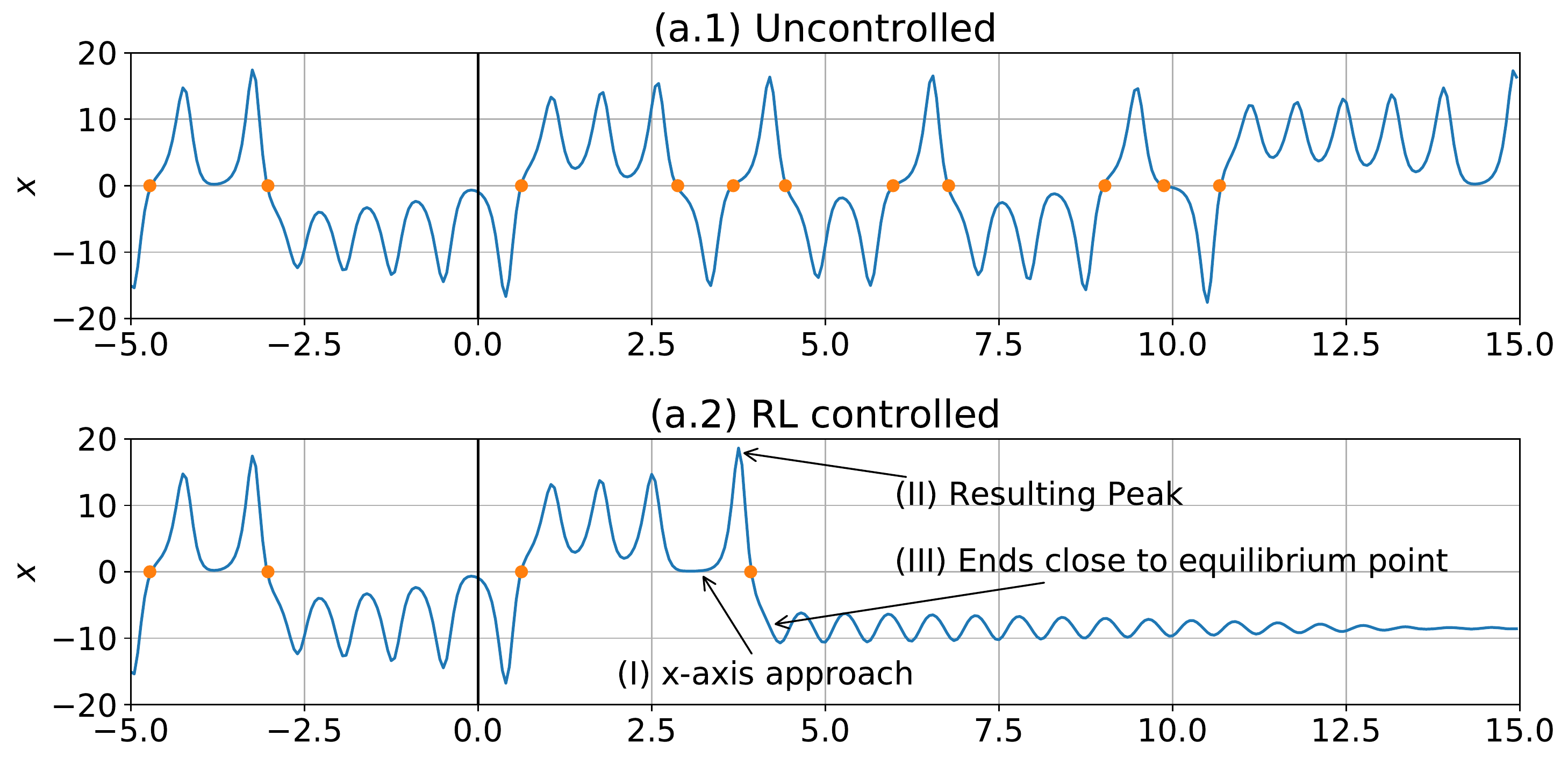}
  \caption{LA stabilizer control problem}
  \label{fig:LA-x-inv}
\end{subfigure}
\begin{subfigure}{1.0\textwidth}
  \centering
  \includegraphics[width=0.66\linewidth]{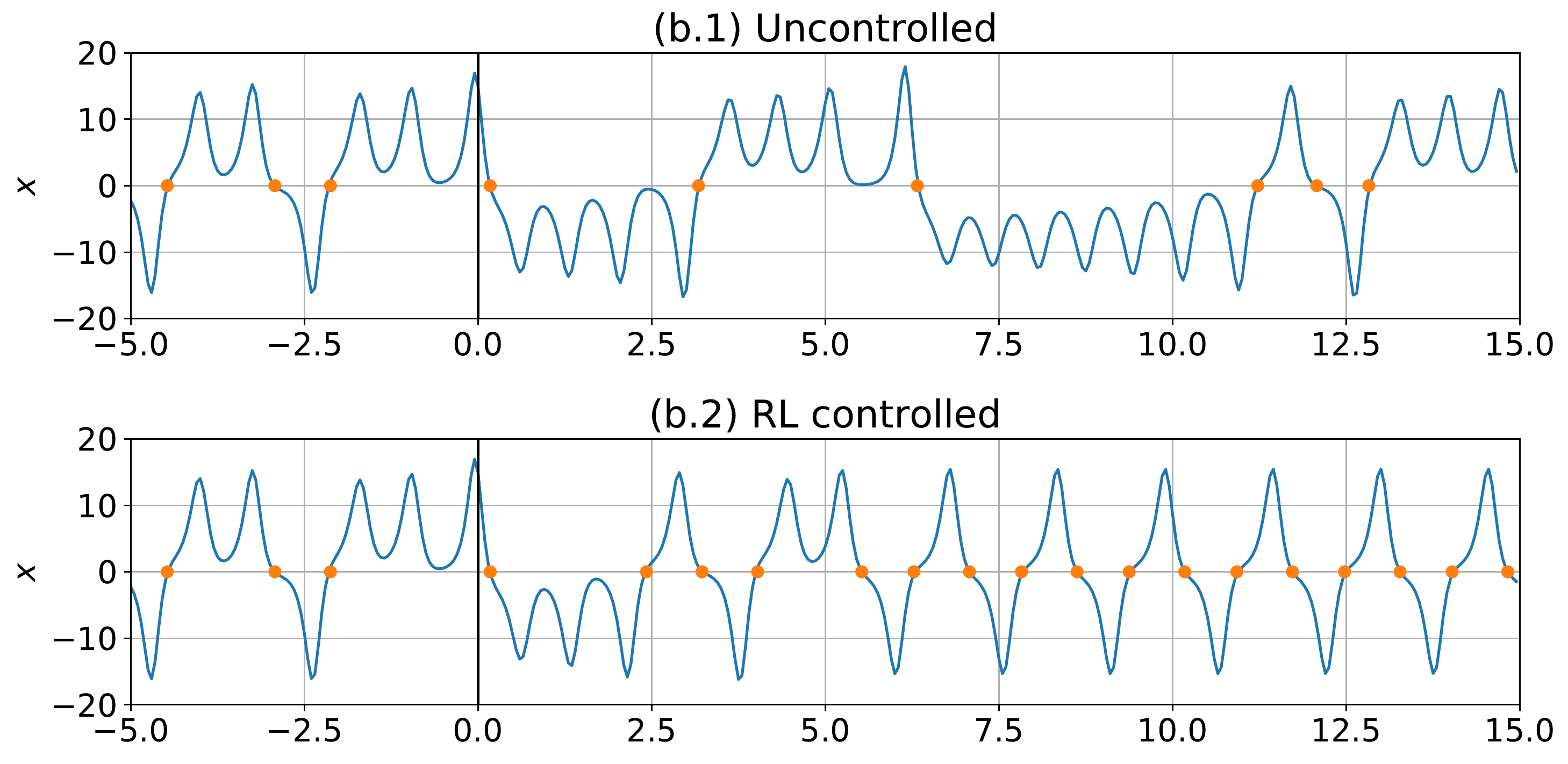}
  \caption{LA oscillator control problem}
  \label{fig:LA-x-nor}
\end{subfigure}
\caption{System trajectories with RL control, respectively aiming at minimizing (a) and maximizing (b) the number of $x$ sign changes. Panel (a) shows that the RL agent is able to fully stabilize the system on an unstable equilibrium by using a complex strategy in three steps (I: controlling the system such that it approaches $x,y,z = 0$ which results in a peak (II) which after going through $x=0$ ends close enough to an unstable equilibrium (III) such that the control is able to fully stabilize the system). Furthermore, Figure~\ref{fig:LA-x-inv} shows that the RL agent is able to find and stabilize a unstable periodic orbits with a desired property of a high frequency of sign changes of x.}
\label{fig:LA-control}
\end{figure}

Applying RL on these control problems with no delay results in the behaviors shown in Figure~\ref{fig:LA-control}. The control  develops complex strategies to maximize/minimize the frequency of sign changes of $x$.

\bibliographystyle{tfnlm}
\bibliography{references}

\end{document}